\documentclass[fleqn]{annalen}
\usepackage{cite}
\usepackage{epsf}
\begin{document}
\newcommand{\volume}{0}              
\newcommand{\xyear}{0000}            
\newcommand{\issue}{0}               
\newcommand{\recdate}{31 December 0000}  
\newcommand{\revdate}{dd.mm.yyyy}    
\newcommand{\revnum}{0}              
\newcommand{\accdate}{1 Januar 0000}    
\newcommand{\coeditor}{bk}           
\newcommand{\firstpage}{1}           
\newcommand{\lastpage}{39}           
\setcounter{page}{\firstpage}        
\newcommand{\keywords}{quantum phase transitions, itinerant magnets,
       metal-insulator transitions}
\newcommand{\PACS}{05.70.FH; 64.60.Ak; 75.45.+j}
\newcommand{\shorttitle}{T. Vojta, Quantum phase transitions in electronic
                         systems}
\title{Quantum phase transitions in electronic systems}
\author{Thomas Vojta}
\newcommand{\address}
{Institut f{\"u}r Physik, Technische Universit\"at Chemnitz, D-09107 Chemnitz,
Germany}
\newcommand{\email}{\tt vojta@physik.tu-chemnitz.de}
\maketitle
\def\tr{{\rm tr}\,}
\def\Tr{{\rm Tr}\,}
\def\sgn{{\rm sgn\,}}
\def\boldphi{\mbox{\boldmath $\phi$}}
\def\boldvarphi{\mbox{\boldmath $\varphi$}}
\def\figuresize{12cm}

\begin{abstract}
Quantum phase transitions occur at zero temperature when some
non-thermal control-parameter like pressure or chemical composition
is changed. They are driven by quantum rather than thermal fluctuations.
In this review we first give a pedagogical introduction to quantum phase
transitions and quantum critical behavior emphasizing similarities with
and differences to classical thermal phase transitions.
We then illustrate the general concepts by discussing a few examples of quantum
phase transitions occurring in electronic systems. The ferromagnetic transition
of itinerant electrons shows a very rich behavior since the magnetization couples
to additional electronic soft modes which generates an effective long-range
interaction between the spin fluctuations. We then consider the influence of rare
regions on quantum phase transitions in systems with quenched disorder, taking
the antiferromagnetic transitions of itinerant electrons as a primary example.
Finally we discuss some aspects of the
metal-insulator transition in the presence of quenched disorder and interactions.
\end{abstract}

\tableofcontents

\section{Classical and quantum phase transitions}
\label{sec:PT}
\subsection{Introduction}

Phase transitions have played, and continue to play,
an essential role in shaping the world. The large scale
structure of the universe is the result of a sequence of
phase transitions during the very early stages
of its development. Later, phase transitions
accompanied the formation of galaxies, stars and planets.
Even our everyday life is unimaginable without the
never ending transformations of water between ice, liquid and
vapor. Understanding phase transitions is a thus
a prime endeavor of physics.

Under normal conditions the phase transitions of water
are so-called first-order transitions. They
involve latent heat, i.e., a finite amount of heat is released
while the material is cooled through an infinitesimally small
temperature interval around the transition temperature.
Phase transitions that do {\em not} involve latent heat, the
so-called continuous transitions, are particularly
interesting since the typical length and time scales of
fluctuations of, e.g., the density, diverge when
approaching the transition point. These divergences and
the resulting singularities of physical observables
are called the critical behavior.
Understanding critical behavior has been a great challenge for theoretical
physics. More than a century has gone by from the first discoveries
until a consistent picture emerged. However, the theoretical concepts
established during this development, viz., scaling and the
renormalization group, now belong to the central paradigms of modern
physics.

The phase transitions we encounter in everyday life occur at finite temperature.
These so-called thermal or classical\footnote{
The justification for calling all thermal phase transitions classical will become
clear in Sec.\ \ref{subsec:qm}}
phase transitions are driven by
thermal fluctuations. In recent years a different class of phase transitions,
the so-called quantum phase transitions, has started to attract a lot of attention.
Quantum phase transitions occur at zero temperature when some
non-thermal control parameter is changed. They are driven by quantum
rather than thermal fluctuations.
Quantum phase transitions in electronic systems
have gained particular attention since some of the most exciting
discoveries in contemporary condensed matter physics,
such as the localization problem, various magnetic
phenomena,
integer and fractional quantum Hall effects, and high-temperature
superconductivity are often attributed to quantum critical points.

The purpose of this review is twofold. The first section gives a
pedagogical introduction to the field of quantum phase transitions
with a particular emphasis on the similarities with and the differences to
classical thermal phase transitions.
After briefly sketching the historical development
the basic concepts of continuous phase
transitions and critical behavior are summarized. We then consider
the question 'How important is quantum mechanics for the physics of phase
transitions?' which leads directly to the distinction between classical
thermal and quantum phase transitions.
In the following sections these ideas are illustrated by discussing
a number of examples of quantum phase transitions occurring in electronic
systems. Specifically, in Sec.\ \ref{sec:QSM} a toy model for a quantum
phase transition is considered, the so-called quantum spherical model. It can
be solved exactly, providing an easily accessible example of a quantum phase
transition. Sec.\ \ref{sec:IQM} contains a discussion of the ferromagnetic
quantum phase transition of itinerant electrons.
It is demonstrated that the coupling of the magnetization to additional
soft modes in the zero-temperature electron system changes the properties
of the transition profoundly. The influence of disorder on quantum phase
transitions is studied in Sec.\ \ref{sec:RR} paying particular attention to
rare disorder fluctuations. It is shown that they can change the universality
class of the transition or even destroy the conventional
critical behavior.
In Sec. \ref{sec:MIT} we discuss some aspects of the metal-insulator
transition of disordered interacting electrons. On the one hand we consider
the influence of local moments on the transition by incorporating them into
a transport theory. On the other hand we study the transition by means of
large-scale numerical simulations. To do this, an efficient numerical method
is developed, called the Hartree-Fock based diagonalization.
It is shown that electron-electron interactions can lead to a considerable
enhancement of transport in the
strongly localized regime.  Finally, Sec.\ \ref{sec:SUM} is devoted to a short summary and
outlook.

\subsection{From critical opalescence to quantum criticality}

In 1869 Andrews \cite{Andrews1869} discovered a very special point in the
phase diagram of carbon dioxide. At a temperature of about 31\,$^\circ$C
and 73 atmospheres pressure the properties of the liquid and the
vapor phases became indistinguishable. In the neighborhood of this point
carbon dioxide strongly scattered light. Andrews called this point
the critical point and the strong light scattering the critical
opalescence. Four years later van der Waals \cite{vanderWaals1873}
presented his
doctoral thesis {'\it On the continuity of the liquid and gaseous
states'} which contained one of the first theoretical explanations of
critical phenomena based on the now famous van der Waals equation
of state. It provides the prototype of a mean-field description of
a phase transition by assuming that the individual interactions
between the molecules
are replaced by an interaction with a hypothetic global
mean field.
In the subsequent years similar behavior was found
for many other materials. In particular, in 1895 Pierre Curie
\cite{Curie1895} noticed
that ferromagnetic iron also shows such a critical point which today is
called the Curie point. It is located at zero magnetic field
and a temperature of
about 770\,$^\circ$C, the highest temperature for which a permanent
magnetization can exist in zero field. At this temperature phases
differing by the direction of the magnetization become obviously
indistinguishable. Again it was only a few years later when
Weiss \cite{Weiss07}
proposed the molecular-field theory of ferromagnetism which
qualitatively explained the experiments. Like the van der Waals theory of the
liquid-gas transition the molecular-field theory of ferromagnetism
is based on the existence of a hypothetic molecular (mean) field.
The so-called classic era of critical phenomena culminated in the
Landau theory of phase transitions \cite{Landau37}. Landau gave some very
powerful and general arguments based on symmetry which suggested
that mean-field theory is essentially exact. While we know today that
this is not the case, Landau theory is still an invaluable starting point
for the investigation of critical phenomena.

The modern era of critical phenomena started when it was realized
that there was a deep problem connected with the values of the
critical exponents which describe how physical quantities vary
close to the critical point. In 1945 Guggenheim
\cite{Guggenheim45} realized that the
coexistence curve of the gas--fluid phase transition is not parabolic,
as predicted by van der Waals' mean-field theory. At about the same time
Onsager \cite{Onsager44} exactly solved the two-dimensional Ising model showing
rigorously that in this system the critical behavior is different from the
predictions of mean-field theory. After these observations it
took about twenty years until a solution of the 'exponent puzzle'
was approached. In 1965 Widom \cite{Widom65}
formulated the scaling hypothesis
according to which the singular part of the
free energy is a generalized homogeneous
function of the parameters. A year later, Kadanoff \cite{Kadanoff66}
proposed a simple
heuristic explanation of scaling based on the argument that at
criticality the system essentially 'looks the same on all length scales'.
The breakthrough came with a series of seminal papers by Wilson \cite{Wilson71}
in 1971. He formalized Kadanoff's heuristic arguments and
developed the renormalization group. For these discoveries, Wilson won the
1982 physics Nobel price.
The development of the renormalization group
initiated an avalanche of activity in the field which still continues.

Today, thermal equilibrium phase transitions are well understood in principle,
even if new interesting
transitions, e.g., in soft condensed matter systems,
continue to be found. In recent years the scientific
interest has shifted towards new fields. One of these fields
are phase transitions in non-equilibrium systems.
They occur, e.g., in systems approaching
equilibrium after a non-infinitesimal perturbation or in systems
driven by external fields or non-thermal
noise to a non-equilibrium (steady) state.
Examples are provided by growing surfaces,
chemical reaction-diffusion systems, or biological systems
(see, e.g., Refs. \cite{Meakin93,MarroDickman97,vanBeijeren98}).
Non-equilibrium phase transitions are characterized
by singularities in the stationary or dynamic
properties of the non-equilibrium states rather than by
thermodynamic singularities.

Another very active avenue of research are quantum phase
transitions which are the topic of this review.
The investigation of quantum phase transitions was pioneered
by Hertz \cite{Hertz76} who built on earlier work by
Suzuki \cite{Suzuki76} and Beal-Monod \cite{Beal-Monod74}.
He developed a renormalization group method
for magnetic transitions of itinerant electrons which was a direct
generalization of Wilson's approach to classical
transitions. He found that
the ferromagnetic transition is mean-field like in all
dimensions $d > 1$. While Hertz' general scaling scenario
at a quantum critical point is valid, his specific predictions for the
ferromagnetic quantum phase transition are incorrect, as will be
explained in Sec.\ \ref{sec:IQM}.

In recent years quantum phase transitions in electronic systems
have attracted considerable attention from theory as well from experiment.
Among the transitions investigated in detail
are the ferromagnetic transition of itinerant electrons, the antiferromagnetic
transition associated with high-temperature superconductivity,
various magnetic transitions in the heavy fermion compounds,
metal-insulator transitions, superconductor-insulator transitions,
and the plateau transition in quantum Hall systems.
This list is certainly incomplete and new transitions continue to be
found.
For reviews on some of these transitions see, e.g., Refs.
\cite{BelitzKirkpatrick94,Sachdev96,SGCS97,
KirkpatrickBelitz97,BelitzKirkpatrick98, SachdevVojta99}.
There is also a very recent text book on quantum phase
transitions by Sachdev \cite{Sachdev99}.

\subsection{Basic concepts of phase transitions and critical behavior}
\label{subsec:concepts}

Since the discoveries of scaling and the renormalization group
a number of excellent text books on
phase transitions and critical behavior have appeared (e.g., those by
Ma \cite{Ma76} or  Goldenfeld \cite{Goldenfeld92}).
Therefore, in this section we only
briefly collect the basic concepts which are necessary for the
later discussion.

A continuous phase transition can usually be characterized by an
order parameter, a concept first introduced by Landau.
An order parameter is a thermodynamic quantity that is zero in one phase
(the disordered) and non-zero and non-unique in the other (the ordered)
phase. Very often the choice of an order parameter for a particular
transition is obvious as, e.g., for the ferromagnetic transition where
the total magnetization is an order parameter. Sometimes, however,
finding an appropriate order parameter is a complicated problem by
itself, e.g., for the disorder-driven localization-delocalization transition
of non-interacting electrons.

While the thermodynamic average of the order parameter is zero in the
disordered phase, its fluctuations are non-zero. If the phase transition
point, i.e., the critical point, is approached the spatial
correlations of the
order parameter fluctuations become long-ranged. Close to the
critical point their typical length scale, the correlation length $\xi$,
diverges as
\begin{equation}
  \xi \propto t^{-\nu}
\end{equation}
where $\nu$ is the correlation length critical exponent and
$t$ is some dimensionless distance from the critical point.
It can be defined as $t=|T-T_c|/T_c$
if the transition occurs at a non-zero temperature $T_c$.
In addition to the long-range correlations in space there are
analogous long-range correlations of the order parameter
fluctuations in time. The typical time scale for a decay of
the fluctuations is the correlation (or equilibration) time
$\tau_c$. As the critical point is approached the correlation
time diverges as
\begin{equation}
  \tau_c \propto \xi^z \propto t^{-\nu z}
\label{eq:correlation time}
\end{equation}
where $z$ is the dynamical critical exponent.
Close to the critical point there is no characteristic
length scale other than $\xi$ and no characteristic time
scale other than $\tau_c$.\footnote{Note that a microscopic
cutoff scale must be present to explain non-trivial critical
behavior, for details see, e.g., Goldenfeld \cite{Goldenfeld92}.
In a solid such a scale is, e.g., the lattice
spacing.}
As already noted by Kadanoff \cite{Kadanoff66}, this is the
physics behind Widom's scaling hypothesis, which we will now
discuss.

Let us consider a classical system, characterized by its Hamiltonian
\begin{equation}
  H(p_i,q_i)= H_{kin}(p_i)+ H_{pot}(q_i)
\end{equation}
where $q_i$ and $p_i$ are the generalized coordinates and momenta,
and $H_{kin}$ and $H_{pot}$ are the kinetic and potential energies,
respectively.\footnote{Velocity dependent potentials like in the case
of charged particles in an electromagnetic field are excluded.}
In such a system 'statics and dynamics decouple', i.e., the momentum
and position sums in the partition function
\begin{equation}
  Z= \int \prod dp_i e^{-H_{kin}/k_B T} ~\int \prod dq_i e^{-H_{pot}/k_B T}
   = Z_{kin}  Z_{pot}
\label{eq:classical Z}
\end{equation}
are completely independent from each other.
The kinetic contribution to the free energy
density $f=-(k_B T/V) \log Z$ will usually not display any singularities,
since it derives from the product of simple Gaussian integrals. Therefore
one can study the critical behavior using effective time-independent
theories like the Landau-Ginzburg-Wilson theory. In this type of theories
the free energy is expressed as a functional of the order parameter
$M({\bf r})$ only. All other degrees of freedom have been integrated
out in the derivation of the theory starting from a microscopic
Hamiltonian. In its simplest form \cite{Landau37,Wilson71,Ginzburg60}
valid, e.g.,
for an Ising ferromagnet, the Landau-Ginzburg-Wilson functional $\Phi[M]$ reads
\begin{eqnarray}
  \Phi[M] &=& \int d^d r ~M({\bf r})
    \left (-\frac {\partial^2}{\partial {\bf r}^2} + t  \right ) M({\bf r})
    + u \int d^d r ~ M^4({\bf r}) - B \int d^d r ~ M({\bf r}), \nonumber \\
      Z &=& \int D[M] e^{-\Phi[M]} \quad ,
\label{eq:classical LGW}
\end{eqnarray}
where $B$ is the field conjugate to the order parameter (the magnetic field
in case of a ferromagnet).

Since close to the critical point the correlation length is the only relevant
length scale, the physical properties must be unchanged, if
we rescale all lengths in the system by a common factor $b$,
and at the same time adjust the external parameters in such a way
that the correlation length retains its old value. This gives rise to
the homogeneity relation for the free energy density,
\begin{equation}
 f(t,B) = b^{-d} f(t\, b^{1/\nu}, B\, b^{y_B}).
\label{eq:widom}
\end{equation}
Here $y_B$ is another critical exponent.
The scale factor $b$ is an arbitrary positive number.
Analogous homogeneity relations for other thermodynamic quantities
can be obtained by differentiating $f$. The homogeneity law
(\ref{eq:widom}) was first obtained phenomenologically by
Widom \cite{Widom65}. Within the framework of the renormalization group
theory it can be derived from first principles.

In addition to the critical exponents $\nu,~ y_B$ and $z$ defined above,
a number of other exponents is in common use. They describe the
dependence of the order parameter and its correlations on the
distance from the
critical point and on the field conjugate to the order parameter.
The definitions of the most commonly used
critical exponents are summarized in Table \ref{table:exponents}.
\begin{table}
\caption{Definitions of the commonly used critical exponents in
  the 'magnetic language', i.e., the order parameter is the magnetization
  $m=\langle M \rangle$, and the conjugate field is a magnetic field $B$. $t$ denotes the
  distance from the critical point and $d$ is the space dimensionality.
  (The exponent $y_B$ defined in (\ref{eq:widom}) is related to $\delta$
  by $y_B=d \, \delta /(1+\delta)$.)}
\renewcommand{\arraystretch}{1.2}
\vspace{2mm}
\begin{tabular*}{\textwidth}{c@{\extracolsep\fill}ccc}
\hline
\hline
&exponent& definition & conditions \\
\hline
specific heat &$\alpha$& $c \propto |t|^{-\alpha}$ & $t \to 0, B=0$\\
order parameter& $\beta$ & $m \propto (-t)^\beta$ & $t \to 0$ from below, $B=0$\\
susceptibility& $\gamma$ & $\chi \propto |t|^{-\gamma}$ & $t \to 0, B=0$\\
critical isotherm & $\delta$ & $B \propto |m|^\delta {\rm sign}(m)$ & $B \to 0, t=0$\\
\hline
correlation length& $\nu$ & $\xi \propto |t|^{-\nu}$ & $t \to 0, B=0$\\
correlation function& $\eta$ & $G(r) \propto |r|^{-d+2-\eta}$ & $t=0, B=0$\\
\hline
dynamical& $z$ & $\tau_c \propto \xi^{z}$ & $t \to 0, B=0$\\
\hline
\hline
\end{tabular*}
\vspace*{2mm}
\label{table:exponents}
\end{table}
Note that not all the exponents defined in Table \ref{table:exponents}
are independent from each other.
The four thermodynamic exponents $\alpha, \beta,\gamma,\delta$ can
all be obtained from the free energy (\ref{eq:widom}) which contains
only two independent exponents.
They are therefore connected by the so-called scaling relations
\begin{eqnarray}
2- \alpha &=&  2 \beta +\gamma~,  \\
2 - \alpha &=& \beta ( \delta + 1)~.
\end{eqnarray}
Analogously, the exponents of the correlation length and correlation
function are connected by two so-called hyperscaling relations
\begin{eqnarray}
2- \alpha &=&  d\,\nu~,  \\
\gamma &=& (2-\eta) \nu~.
\end{eqnarray}
Since statics and dynamics decouple in classical statistics the dynamical
exponent $z$ is completely independent from all the others.

The critical behavior at a particular phase transition is
completely characterized by the set of critical exponents. One of the most
remarkable features of continuous phase transitions is universality, i.e.,
the fact that the critical exponents are the same for entire classes of
phase transitions which may occur in very different physical systems.
These classes, the so-called universality classes, are determined only
by the symmetries of the Hamiltonian and the spatial dimensionality
of the system. This implies that the critical exponents of a
phase transition occurring in nature can be determined exactly
(at least in principle) by investigating
any simplistic model system belonging to the same universality class,
a fact that makes the field very attractive for theoretical physicists.
The mechanism behind universality is again the divergence of the correlation
length. Close to the critical point the system effectively averages over
large volumes rendering the microscopic details of the Hamiltonian
unimportant.

The critical behavior at a particular transition is crucially determined
by the relevance or irrelevance of order parameter fluctuations.
It turns out that fluctuations become increasingly important if the
spatial dimensionality of the system is reduced. Above a certain
dimension, called the upper critical dimension $d_c^+$, fluctuations are
irrelevant, and the critical behavior is identical to that predicted by
mean-field theory (for systems with short-range interactions and a scalar
or vector order parameter $d_c^+=4$). Between $d_c^+$ and a second special dimension, called
the lower critical dimension $d_c^-$, a phase transition still exists but
the critical behavior is different from mean-field theory. Below $d_c^-$
fluctuations become so strong that they completely suppress the ordered phase.

\subsection{How important is quantum mechanics?}
\label{subsec:qm}

The question of to what extent quantum mechanics is important
for understanding a continuous phase transition is a multi-layered
question. One may ask, e.g., whether quantum mechanics is necessary
to explain the existence and the properties of the ordered phase.
This question can only be decided on a case-by-case basis, and
very often quantum mechanics is essential as, e.g., for superconductors.
A different question to ask would be how important quantum mechanics
is for the asymptotic behavior close to the critical point and thus for
the determination of the universality class the transition belongs to.

It turns out that the latter question has a remarkably clear and
simple answer: Quantum mechanics does {\em not} play any role for
the critical behavior if the transition occurs at a finite temperature.
It does play a role, however, at zero temperature. In the following
we will first give a simple argument explaining these facts.
To do so it is useful to distinguish fluctuations with predominantly thermal
and quantum character depending on whether their thermal energy $k_B T$ is
larger or smaller than the quantum energy scale $\hbar \omega_c$.
We have seen in the preceeding section that the typical time scale $\tau_c$
of the fluctuations diverges as a continuous transition is approached.
Correspondingly, the typical frequency scale $\omega_c$ goes to zero and with it the typical
energy scale
\begin{equation}
  \hbar \omega_c \propto |t|^{\nu z}~.
  \label{eq:energy scale}
\end{equation}
Quantum fluctuations will be important as long as this typical energy scale
is larger than  the thermal energy $k_B T$.
If the transition occurs
at some finite temperature $T_c$ quantum mechanics will thus become
unimportant for $|t|<t_x$ with the crossover distance $t_x$ given by
\begin{equation}
 t_x \propto T_c^{1/\nu z}~.
 \label{eq:crossover t}
\end{equation}
We thus find that the
critical behavior asymptotically close to the transition is entirely
classical if the transition temperature $T_c$ is nonzero. This justifies
to call all finite-temperature phase transitions classical transitions,
even if the properties of the ordered state are completely determined
by quantum mechanics as is the case, e.g., for the superconducting
phase transition of, say, mercury at $T_c=4.2$ K.
In these cases quantum fluctuations are obviously
important on microscopic scales, while classical thermal fluctuations
dominate on the macroscopic scales that control the critical behavior.
If, however, the transition occurs at zero temperature as a function of
a non-thermal parameter like the pressure $p$, the crossover distance $t_x=0$.
(Note that at zero temperature the distance $t$ from the critical point
cannot be defined via the reduced temperature. Instead, one can define
$t=|p-p_c|/p_c$.)
Thus, at zero temperature the condition
$|t|<t_x$ is never fulfilled, and quantum mechanics
will be important for the critical behavior. Consequently,
transitions at zero temperature are called quantum phase transitions.

Let us now generalize the homogeneity law (\ref{eq:widom}) to the
case of a quantum phase transition. We consider a system
characterized by a Hamiltonian $H$.
In a quantum problem kinetic and potential part of $H$
in general do not commute. In contrast to the classical partition
function (\ref{eq:classical Z}) the quantum mechanical partition function
does {\em not} factorize, i.e., 'statics and dynamics are always coupled'.
The canonical density operator $e^{-H/k_B T}$ looks
exactly like a time evolution operator in imaginary time
$\tau$ if one identifies
\begin{equation}
1/k_B T = \tau = -i\Theta /\hbar
\label{eq:imaginary time}
\end{equation}
where $\Theta$ denotes the real time. This naturally
leads to the introduction of an imaginary time direction into the
system. An order parameter field theory analogous to
the classical Landau-Ginzburg-Wilson theory (\ref{eq:classical LGW})
therefore needs to be formulated in terms of space and time dependent fields.
The simplest example of a quantum Landau-Ginzburg-Wilson functional,
valid for, e.g., an Ising model in a transverse field, reads
\begin{eqnarray}
  \Phi[M] &=& \int_0^{1/k_B T} d\tau \int d^d r ~M({\bf r},\tau)
    \left (-\frac {\partial^2}{\partial {\bf r}^2}
    -\frac {\partial^2}{\partial \tau^2}+ t  \right ) M({\bf r},\tau)
    ~+ \nonumber\\
    &+& u \int_0^{1/k_B T} d\tau \int d^d r ~ M^4({\bf r},\tau)
    ~-~ B \int_0^{1/k_B T} d\tau \int d^d r ~ M({\bf r},\tau)~.
      \label{eq:quantum LGW}
\end{eqnarray}
Let us note that the coupling of statics and dynamics
in quantum statistical dynamics also leads to the fact that the
universality classes for quantum phase transitions are smaller
than those for classical transitions. Systems which belong
to the same classical universality class may display
different quantum critical behavior, if their dynamics differ.

The classical homogeneity law (\ref{eq:widom}) for the free energy density can
now easily be adopted to the case of a quantum phase transition.
At zero temperature the imaginary time acts similarly to an additional
spatial dimension since the extension of the system in this direction
is infinite. According to (\ref{eq:correlation time}), time scales like
the $z$th power of a length. (In the simple example
(\ref{eq:quantum LGW}) space and time enter the
theory symmetrically leading to $z=1$.)
Therefore, the homogeneity law for the free energy density at
zero temperature reads
\begin{equation}
 f(t,B) = b^{-(d+z)} f(t\, b^{1/\nu},B\, b^{y_B})~.
\label{eq:quantum widom}
\end{equation}
Comparing (\ref{eq:quantum widom}) and (\ref{eq:widom})
directly shows that a quantum phase transition in $d$ dimensions is
equivalent to a classical transition in $d+z$ spatial dimensions.
Thus, for a quantum phase transition
the upper critical dimension, above which
mean-field critical behavior becomes exact, is reduced by $z$
compared to the corresponding classical transition.

\begin{figure}[t]
  \epsfxsize=\figuresize
  \centerline{\epsffile{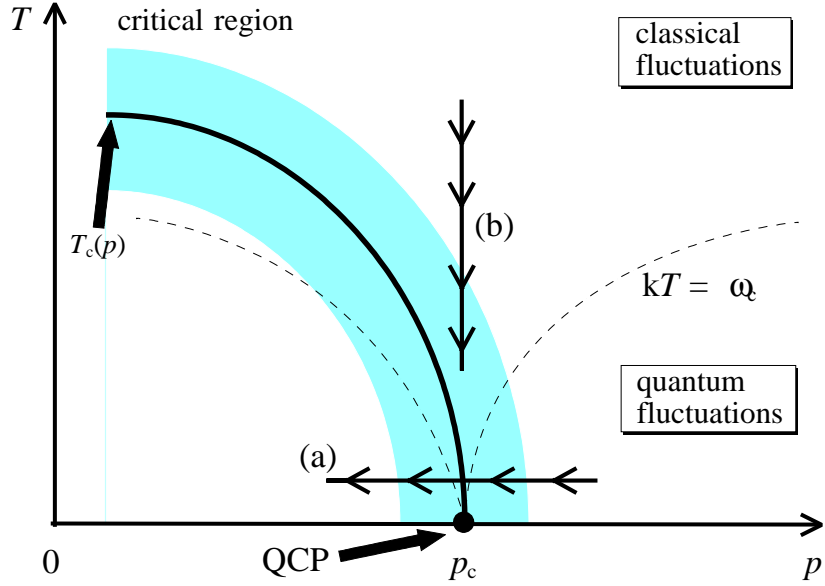}}
  \caption{Schematic phase diagram in the vicinity of a quantum
     critical point (QCP). The solid line marks the boundary between ordered
     and disordered phase. The dashed lines indicate the crossover
     between predominantly quantum or classical character of the fluctuations,
     and the shaded area denotes the critical region where the leading
     critical singularities can be observed. Paths (a) and (b) are discussed
     in the text.
     }
\label{fig:schematic phase diagram}
\end{figure}

Now the attentive reader may ask: Why are quantum phase transitions more than an academic
problem? Any experiment is done at a non-zero temperature where,
as we have explained above, the asymptotic critical behavior is classical.
The answer is again provided by the crossover condition
(\ref{eq:crossover t}): If the transition
temperature $T_c$ is very small quantum fluctuations will remain important
down to very small $t$, i.e., very close to the transition.
At a more technical level, the behavior at small but non-zero temperatures
is determined by the crossover between two types of critical behavior,
viz. quantum critical behavior at $T=0$ and classical critical behavior
at non-zero temperatures. Since the 'extension of the system in
imaginary time direction' is given by the inverse temperature $1/k_B T$ the
corresponding crossover scaling is equivalent to finite size scaling in
imaginary time direction. The crossover from quantum to classical behavior will
occur when the correlation time $\tau_c$ reaches $1/k_B T$ which is equivalent
to the condition (\ref{eq:crossover t}).
By adding the temperature as an explicit parameter and taking into
account that in imaginary-time formalism it scales like an
inverse time (\ref{eq:imaginary time}), we can generalize the quantum
homogeneity law (\ref{eq:quantum widom}) to finite temperatures,
\begin{equation}
 f(t,B,T) = b^{-(d+z)} f(t\, b^{1/\nu},B\, b^{y_B}, T \, b^z)~.
\label{eq:temperature quantum widom}
\end{equation}
The resulting phase diagram close to a quantum critical points will
be of one of two qualitative different types.
The first type describes situations where an ordered phase exists at finite
temperature. These phase diagrams are illustrated in Fig.\
\ref{fig:schematic phase diagram}.
Here $p$ stands for the (non-thermal) parameter which tunes the quantum phase
transition. According to (\ref{eq:crossover t}) the vicinity of the quantum
critical point can be divided into regions with predominantly classical
or quantum fluctuations. The boundary, marked by the dashed lines in Fig.\
\ref{fig:schematic phase diagram}, is not sharp
but rather a smooth crossover line. At sufficiently low temperatures
these crossover lines are inside the critical region (i.e., the region
where the leading critical power laws can be observed).
An experiment performed along path (a) will therefore observe a
crossover from quantum critical behavior away from the transition
to classical critical behavior asymptotically close to it. At very low
temperatures the classical region may become so narrow that it is actually
unobservable in an experiment.

In addition to the critical behavior at very low temperatures,
the quantum critical point also controls
the behavior in the so-called quantum critical region \cite{CHN89}.
This region is located at the critical $p$ but, somewhat
counter-intuitively, at comparatively high temperatures
(where the character of the fluctuations is classical). In this region
the system 'looks critical' with respect to $p$ but is driven away from
criticality by the temperature (i.e., the critical singularities are
exclusively protected by $T$).
An experiment carried out along path (b) will therefore observe the
temperature scaling at the quantum critical point.

The second type of phase diagram occurs
if an ordered phase exists at zero temperature only (as is the case for
two-dimensional quantum antiferromagnets). In this case there will be no
true phase transition in any experiment.  However, the system will display
quantum critical behavior in the above-mentioned quantum critical region
close to the critical $p$.

\section{Quantum spherical model}
\label{sec:QSM}

\subsection{Classical spherical model}

In the process of understanding a novel physical problem it is
often very useful to consider a simple model which displays the
phenomena in question in their most basic form.
In the field of classical equilibrium critical phenomena
such a model is the so-called classical spherical model which
is one of the very few models in statistical physics that
can be solved exactly but
show non-trivial (i.e., non mean-field) critical behavior.
The spherical model was
conceived by Kac in 1947 in an attempt to simplify the Ising model.
The basic idea was to replace the discrete Ising spins having
only the two possible values $S_i=\pm 1$ by continuous real variables
between $-\infty$ and $\infty$ so that the multiple sum in the
partition function of the Ising model is replaced by a multiple
integral which should be easier to perform.
However, the multiple integral turned out to be not at all simple,
and for a time it looked as if the spherical model was actually
harder to solve than the corresponding Ising model. Eventually Berlin and
Kac \cite{BerlinKac52}
solved the spherical model by using the method of steepest descent to perform
the integrals over the spin variables. Stanley \cite{Stanley68} showed that
the spherical model, though created to be a simplification of the
Ising model, is equivalent to the $n \to \infty$ limit of the classical
$n$-vector model.\footnote{In the classical $n$-vector model the
dynamical variables are $n$-dimensional unit vectors. Thus, the Ising model
is the 1-vector model, the classical XY-model is the 2-vector model and
the classical Heisenberg model is the 3-vector model.} Therefore, it can be
used as the starting point for a $1/n$-expansion of the critical behavior.

In the following years the classical spherical model was solved exactly not
only for nearest neighbor ferromagnetic interactions but also for
long-range power-law interactions \cite{Joyce66}, random interactions
\cite{KTJ76,JagannathanRudnick89}, systems in random magnetic fields
\cite{HornreichSchuster82,Vojta93},
and disordered electronic systems with localized states
\cite{VojtaSchreiber94}.
Moreover, the model has been used as a test
case for the finite-size scaling hypothesis
\cite{FisherBarber72,Rudnick90}. Reviews on the classical
spherical model were given by Joyce \cite{Joyce72} and Khorunzhy et al.
\cite{KKPS92}.

Because the classical spherical model possesses such a wide variety of
applications in the field of classical critical phenomena,
it seems natural to look for a quantum version of the model in order
to obtain a toy model for quantum critical behavior. Actually, the history of
quantum spherical models dates back at least as far as the
history of quantum critical behavior. In 1972 Obermair \cite{Obermair72}
suggested
a canonical quantization scheme for a dynamical spherical model.
However, this and later studies focused on the classical finite temperature
critical behavior of the quantum model and did not consider the
properties of the zero temperature quantum phase transition.

\subsection{Quantization of the spherical model}

The classical spherical model consists of $N$ real variables
$S_i \in (-\infty, \infty)$ that interact with an external
field $h_i$ and with each other via a pair potential
$U_{ij}$. The Hamiltonian is given by
\begin{equation}
 H_{cl}= \frac 1 2 \sum _{i,j}  U_{ij} S_i S_j + \sum_i h_i S_i~.
\label{eq:Hspcl}
\end{equation}
In order to make the model well-defined at low temperatures, i.e.,
in order to prevent a divergence of $S_i$ in the ordered phase,
the values of $S_i$ are subject to an additional constraint, the
so-called spherical constraint. Two versions of the constraint
have been used in the literature, the strict and the mean constraints,
defined by
\begin{eqnarray}
\sum_i S_i^2 &=& N,\\
\sum_i \langle S_i^2 \rangle &=& N,
\label{eq:sphcon}
\end{eqnarray}
respectively. Here $\langle \ldots \rangle$ is the thermodynamic
average. Both constraints have been shown to
give rise to the same thermodynamic behavior while other
quantities like correlation functions differ. In the following
we restrict ourselves to the mean spherical constraint which is
easier to implement in the quantum case.
The Hamiltonian (\ref{eq:Hspcl}) has no internal dynamics.
According to the factorization (\ref{eq:classical Z}) it can be
interpreted as being only the configurational part of a more complicated
problem. Therefore, the construction of the quantum model consists
of two steps: First we have to add an appropriate kinetic
energy to the Hamiltonian which defines a dynamical spherical model
which can be quantized in a second step.

In order to construct the kinetic energy term we define
canonically conjugate momentum variables $P_i$ which fulfill
the Poisson bracket relations $\{S_i,P_j\}=\delta_{ij}$. The simplest choice
of a kinetic energy term is then the one of Obermair \cite{Obermair72},
$H_{kin}= \frac g 2 \sum_i P_i^2$,
where $g$ can be interpreted as inverse mass. In this case, the
complete Hamiltonian of the dynamical spherical model
\begin{equation}
H= H_{kin} + H_{cl} = \frac g 2 \sum_i P_i^2 + \frac 1 2 \sum_{i,j}
U_{ij} S_i S_j + \sum_i h_i S_i + \mu \left ( \sum_i S_i^2 -N \right )
\label{eq:Hspqu}
\end{equation}
is that of a system of coupled harmonic oscillators. Here we have also
added a source term for the mean spherical constraint (\ref{eq:sphcon}).
(The value of $\mu$ has to be determined self-consistently so that
(\ref{eq:sphcon}) is fulfilled.)

In order to quantize the dynamical spherical model (\ref{eq:Hspqu})
we use the usual canonical quantization scheme: The variables $S_i$
and $P_i$ are reinterpreted as operators. The Poisson bracket relations
are replaced by the corresponding canonical commutation relations
\begin{equation}
[S_i,S_j]=0, ~[P_i,P_j]=0, ~{\rm and}~ [S_i,P_j]=i \hbar \delta_{ij}~.
\label{eq:commu}
\end{equation}
Equations (\ref{eq:sphcon}), (\ref{eq:Hspqu}), and (\ref{eq:commu})
completely define the quantum spherical model. At large $T$ or
$g$ the model is in its disordered phase $\langle S_i \rangle=0$.
The transition to an ordered state can be triggered by lowering
$g$ and/or $T$.

It must be emphasized that
this model does not mimic (or even describe) Heisen\-berg-Dirac spins.
Instead it is equivalent to the $n \to \infty$ limit of a quantum rotor
model which can be seen as a generalization of an Ising model in a
transverse field. Of course, the choices of the kinetic energy and
quantization scheme are not unique.  In agreement with the general
discussion in Sec.\ \ref{subsec:qm} different choices will
lead to different critical behavior at the quantum phase transition,
while the classical critical behavior is the same for all these models.
An example of a different quantization of the spherical model was given
by Nieuwenhuizen \cite{Nieuwenhuizen95}. It leads to a dynamical behavior
that more closely resembles that of Heisenberg-Dirac spins than
our choice. For a more detailed discussion of these questions see
also Ref. \cite{Vojta96}.

\subsection{Quantum phase transitions}

The quantum spherical model defined in eqs.
(\ref{eq:sphcon}), (\ref{eq:Hspqu}), and (\ref{eq:commu})
can be solved exactly since it is equivalent to
a system of coupled harmonic oscillators. This was done
in Ref.\ \cite{Vojta96} for a model with arbitrary translationally invariant
interactions (long-range as well as short-range) in a
spatially homogeneous external field.
The resulting free energy reads
\begin{equation}
f = - \frac {k_B T} N \ln Z = -\mu - \frac {h^2} {4 \mu} + \frac {k_B T} N \sum_k
    \ln \left( 2 \sinh \frac {\omega(k)}{2k_B T} \right )~,
\end{equation}
with $\omega(k)$ given by $\omega^2 = 2 g [\mu +U(k)/2]$, where $U(k)$ is the
Fourier transform of the interaction $U_{ij}$.
The spherical constraint which determines $\mu$ is given by
\begin{equation}
0 =\frac {\partial f} {\partial \mu} = -1 + \frac {h^2} {4 \mu^2} + \frac {1} N \sum_k
    \frac g {2 \omega(k)}  \coth \frac {\omega(k)}{2k_B T}~.
\label{eq:constraint}
\end{equation}
As usual in spherical models the critical behavior is determined by
the properties of the solutions of (\ref{eq:constraint}) for small $\mu$.
At any finite temperature
the $\coth$-term can be expanded giving the same leading long-wavelength
and low frequency terms as in the
classical spherical model (\ref{eq:Hspcl}).
As expected, the resulting critical behavior
at finite temperatures is therefore that of the classical spherical  model.

At zero temperature, the $\coth$-term in (\ref{eq:constraint}) is identical to
one. Thus, the leading long-wavelength and low frequency terms are
different from the classical case. This gives rise to the quantum critical behavior
being different from the classical one.
If the interaction $U_{ij}$ in the Hamiltonian
is short ranged, the dynamical exponent turns out to be $z=1$.
For a power-law interaction, parameterized by the
singularity of the Fourier transform of the interaction,
$U_k \propto |k|^x$ for $k \to 0$, we obtain $z=x/2$.
In both cases the quantum
critical behavior of the $d$-dimensional quantum spherical model
is the same
as the classical critical behavior of a
corresponding $d+z$-dimensional model.
The critical exponents for the quantum and classical phase transitions
are summarized in Table \ref{table:QSM}.
\begin{table}
\caption{Critical exponents at the quantum and classical phase transitions
   of the quantum spherical model as functions of the dimensionality $d$ and
   the exponent $x$ which characterizes the long-wavelength behavior of the
   interaction $U(k) \sim |k|^x$ (short-range interactions correspond to
   $x=2$).}
\renewcommand{\arraystretch}{1.2}
\vspace{2mm}
\begin{tabular*}{\textwidth}{c@{\extracolsep\fill}cccc}
\hline
\hline
        & Quantum transition & Classical transition &  Both\\
exponent& $d<d_c^+=3x/2$ & $d<d_c^+=2x$ &$d>d_c^+$\\
\hline
$\alpha$& $(2d - 3x)/(2d-x)$ & $(d-2x)/(d-x)$ & 0   \\
$\beta$ & 1/2                & 1/2            & 1/2 \\
$\gamma$& $2x/(2d-x)$        & $x/(d-x)$      & 1   \\
$\delta$& $(2d+3x)/(2d-x)$   & $(d+x)/(d-x)$  & 3   \\
$\nu$   & $2/(2d-x)$         & $1/(d-x)$      & $1/x$\\
$\eta$  & $2-x$              & $2-x$          & $2-x$\\
$z$     & $x/2$              & $x/2$          & $x/2$\\
\hline
\hline
\end{tabular*}
\vspace*{2mm}
\label{table:QSM}
\end{table}

In order to describe the crossover between the quantum and
classical critical behaviors the crossover scaling
form of the equation of state was derived. This is only possible below the
upper critical dimension. Above, crossover scaling
breaks down. This is analogous to the breakdown of finite-size
scaling in the spherical model above the upper critical dimension.
It can be attributed to a dangerous irrelevant variable.

In Ref.\ \cite{VojtaSchreiber96}
the influence of a quenched random field on the quantum
phase transition was  considered.
The quantum spherical model can be solved exactly even in the
presence of a random field without the necessity to use the
replica trick. It was found that the quantum critical behavior
is dominated by the static random field fluctuations rather
than by the quantum fluctuations. Since the random field
fluctuations are identical at zero and finite temperatures
it follows that in the presence of a random field quantum and
classical critical behavior are identical.

\section{Ferromagnetic quantum phase transition of itinerant electrons}
\label{sec:IQM}

\subsection{Itinerant ferromagnets}

In the normal metallic state the electrons form a Fermi liquid, a concept
introduced by Landau \cite{Landau56,Landau57}.
In this state the excitation spectrum is
very similar to that of a non-interacting Fermi gas. The basic excitations
are weakly interacting fermionic quasiparticles which behave like normal
electrons but have renormalized parameters like an effective mass.
However, at low temperatures the Fermi liquid is potentially unstable against
sufficiently strong interactions, and
some type of a symmetry-broken state may form.
This low-temperature phase may be a superconductor, a charge density
wave, or a magnetic phase, e.g., a ferromagnet, an anti-ferromagnet,
or a spin glass, to name a few possibilities.
In general, it will depend on the microscopic parameters
of the material under consideration what the nature of the
low-temperature phase and, specifically, of the ground state will be.
Upon changing these microscopic parameters at zero temperature, e.g.,
by applying pressure or an external field or by changing chemical
composition, the nature of the ground state may change, i.e.,
the system may undergo a quantum phase transition.

In this Section we will discuss a particular example of such a
quantum phase transition, viz. the ferromagnetic quantum phase
transition of itinerant electrons.
Most of the Section will be devoted to clean itinerant electrons
but we will
also briefly consider the influence of disorder on the ferromagnetic
transition.

The experimentally best studied example of a ferromagnetic
quantum phase transition of itinerant electrons is probably
provided by the pressure-tuned transition in MnSi
\cite{PML95,PMJL97}. MnSi belongs to the class of so-called
nearly or weakly ferromagnetic materials. This group of metals,
consisting of transition metals and their compounds such as ZrZn$_2$,
TiBe$_2$, Ni$_3$Al, and YCo$_2$ in addition to MnSi are
characterized by strongly enhanced spin fluctuations. Thus, their
ground state is close to a ferromagnetic instability which makes them good
candidates for actually reaching the ferromagnetic quantum phase  transition
in experiment by changing the chemical composition or applying pressure.

\begin{figure}[t]
  \epsfxsize=8.75cm
  \epsfclipon
  \centerline{\epsffile{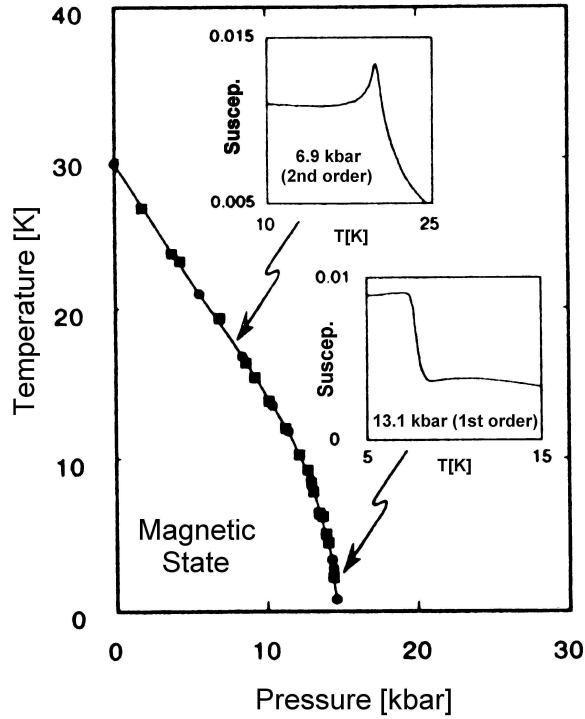}}
  \epsfclipoff
  \caption{Phase diagram of MnSi. The insets show the behavior of
   the susceptibility close to the transition. (after Ref.\
   \protect\cite{PML95}) }
  \label{fig:MnSi}
\end{figure}

At ambient pressure MnSi is paramagnetic for temperatures larger
than $T_c=30$\,K. Below $T_c$ it orders magnetically. The order is,
however, not exactly ferromagnetic but a long-wavelength
(190\,\AA)  helical spin spiral along the (111) direction of the
crystal. The ordering wavelength depends only weakly on
the temperature, but a homogeneous magnetic field of about 0.6\,T suppresses
the spiral and leads to ferromagnetic order.
One of the most remarkable findings about the magnetic phase transition
in MnSi is that it changes from continuous to first order with decreasing
temperature as is shown in Fig.\ \ref{fig:MnSi}.
Specifically, in an experiment carried out at low
pressure (corresponding to a comparatively high transition
temperature) the susceptibility shows a pronounced maximum
at the transition, reminiscent of the singularity expected
from a continuous phase transition.
In contrast, in an experiment at a pressure very close to (but still smaller
than) the critical pressure the susceptibility does not show any sign of a
divergence at the phase transition. Instead, it displays a finite step
suggestive of a first-order phase transition.

A related set of experiments is devoted to a phenomenon called the
{\em itinerant electron metamagnetism}. Here a high magnetic field is
applied to a nearly ferromagnetic material such as
Co(Se$_{1-x}$S$_x$)$_2$ \cite{Adachi79} or Y(Co$_{1-x}$Al$_x$)$_2$
\cite{SGYF90}.  At a certain field strength the magnetization
of the sample shows a pronounced jump. This can easily be explained
if we assume that the free energy
as a function of the magnetization has the triple-well structure
characteristic of the vicinity of a first-order phase transition.
In zero field the side minima must have a larger free energy than the
center minimum (since the material is paramagnetic in zero field).
The magnetic
field essentially just ''tilts'' the free energy function. If one of the side
minima becomes lower than the center (paramagnetic) one, the magnetization jumps.

In the literature the first-order transition in MnSi at low temperatures
as well
as the itinerant electron
meta\-magnetism have been attributed to sharp structures in the electronic
density of states close to the Fermi energy which stem from the
band structure of the particular material. These
structures in the density of states can lead to a negative
quartic coefficient in a magnetic Landau theory and thus to the
above mentioned triple-well structure. In the next section it will
be shown, however, that the two phenomena are generic since they are rooted in
the universal many-body physics underlying the transition.
Therefore, they will occur
for all nearly or weakly  ferromagnetic materials irrespective of
special structures in the density of states.

\subsection{Landau-Ginzburg-Wilson theory of the ferromagnetic
    quantum phase transition}

From a theoretical point of view, the ferromagnetic transition of
itinerant electrons is one of the most obvious quantum phase transitions.
It was also one of the first quantum phase transitions investigated
in some detail. Hertz \cite{Hertz76} studied a simple microscopic model
of interacting electrons and derived a Landau-Ginzburg-Wilson theory
for the ferromagnetic quantum phase transition. Hertz then analyzed
this theory by means of renormalization group methods which were
a direct generalization of Wilson's treatment of classical
transitions. He found a dynamical exponent of $z=3$. According to the
discussion in Sec.\ \ref{subsec:qm} this effectively increases
the dimensionality of the system from $d$ to $d+3$. Therefore,
the upper critical dimension of the quantum phase transition
would be $d_c^+=1$, and Hertz concluded that the critical behavior
of the ferromagnetic quantum phase transition is mean-field like
in all physical dimensions $d>1$. While it was later found \cite{Millis93}
that Hertz' description of the finite temperature phenomena close
to the quantum critical point was incomplete, it was
generally believed that the main qualitative results of his model
at zero temperatures apply to real itinerant ferromagnets as
well.\footnote{In order to obtain a quantitative description
Moriya and Kawabata developed a more sophisticated theory,
the so-called self-consistent renormalization theory of spin fluctuations
\cite{Moriya85}. This theory is very successful in describing magnetic
materials with strong spin fluctuations outside the critical region.
Its results for the critical
behavior at the ferromagnetic quantum phase transition are, however,
identical to those of Hertz.} However, in 1994 Sachdev \cite{Sachdev94}
showed that Hertz' results in dimensions below one (an academic but still
interesting case) violate an exact exponent equality.

Vojta, Belitz, Narayanan, and Kirkpatrick \cite{VBNK96}
have revisited the ferromagnetic transition of itinerant electrons.
They have shown that the properties of the transition are
much more complicated since the magnetization
couples to additional, non-critical soft modes in the electronic system.
Mathematically, this renders the conventional Landau-Ginzburg-Wilson approach
invalid since an expansion of the free energy in powers of the order
parameter does not exist. Physically, the additional soft modes
lead to an effective
long-range interaction between the order parameter fluctuations.
This long-range interaction, in turn, can change the character
of the transition from a
continuous transition with mean-field exponents to either
a continuous transition with non-trivial (non mean-field)
critical behavior or even to a first
order transition.

The derivation of the order parameter field theory
\cite{VBNK96,VBNK97} follows Hertz \cite{Hertz76}
in
spirit, but the technical details are considerably different.
Let us consider a microscopic model Hamiltonian
$H=H_0 + H_{\rm ex}$ of interacting fermions.  $H_{\rm ex}$
 is the exchange
interaction which is responsible for the ferromagnetism,
$H_0$ does not only contain the free electron part
but also all interactions except for the exchange
interaction.
Using standard manipulations (see, e.g., Ref. \cite{NegeleOrland88})
the partition function is written in terms
of a functional integral over fermionic (Grassmann) variables.
After introducing the magnetization field ${\bf M}({\bf r},\tau)$
via a Hubbard-Stratonovich transformation \cite{Hubbard59,Stratonovich57}
of the exchange interaction, a cumulant expansion is used to
integrate out the fermionic degrees of freedom.
The partition function $Z$ takes the form
\begin{equation}
Z = e^{-F_0/T} \int D[{\bf M}]\,\exp\bigl[-\Phi[{\bf M}]\bigr]\quad,
\label{eq:7a}
\end{equation}
where $F_0$ is the non-critical part of the free energy.
With the four-vector notation with $x = ({\bf x},\tau)$
and $\int dx = \int d{\bf x} \int_0^{1/k_B T} d\tau$
the resulting Landau-Ginzburg-Wilson free energy functional reads
\begin{eqnarray}
\Phi[{\bf M}] &=& {1\over 2} \int dx\,dy\
  \left[{1\over \Gamma_t}\delta(x-y) - \chi^{(2)}(x-y)\right]
  {\bf M}(x)\cdot {\bf M}(y) +
\label{eq:Hertz-LGW}  \\
&+& \sum_{n=3}^{\infty}\frac {(-1)^{n+1}}{n!} \int dx_1 \ldots dx_n\
  \chi^{(n)}_{a_1\ldots a_n} (x_1, \ldots, x_n)
  M^{a_1}(x_1)\ldots M^{a_n}(x_n) \nonumber
\end{eqnarray}
where $\Gamma_t$ is the spin-triplet (exchange) interaction
strength.
The coefficients in the Landau-Ginzburg-Wilson functional are the connected
$n$-point spin density correlation functions
$\chi^{(n)}$ of the
reference system $H_0$ which is a conventional Fermi
liquid. The famous Stoner criterion \cite{Stoner38}
of ferromagnetism, $\Gamma_t\, g(\epsilon_F) > 1$
(here $g(\epsilon_F)$ is the density of states at the Fermi energy)
can be rediscovered from
the stability condition of the Gaussian term of $\Phi[{\bf M}]$, if one takes
the spin susceptibility $\chi^{(2)}$ to be that of non-interacting electrons
(in which case $\chi^{(2)}({\bf q}\to 0, \Omega=0) = g(\epsilon_F)$).

The long-wavelength and long-time properties
of the spin-density correlation functions of a Fermi liquid
were studied \cite{BKV97} using
diagrammatical perturbation theory in the interaction.
Somewhat surprisingly, all these correlation functions
generically (i.e., away from any critical point)
show long-range correlations in real space which correspond
to singularities in momentum space in the long-wavelength limit
${\bf q} \to 0$.
While analogous generic long-range correlations in time
(the so-called long-time tails) are well known from
several interacting systems, long-range {\em spatial} correlations
in classical systems are impossible due to the
fluctuation-dissipation theorem. They are known, however,
in non-equilibrium steady states (see, e.g., Ref.\
\cite{SchmittmannZia98}).
The physical reason for the singularities in the coefficients
$\chi^{(n)}$ of the Landau-Ginzburg-Wilson functional is that
in the process of integrating out the fermionic degrees of freedom
the {\em soft} particle-hole excitations  have been integrated
out, too. It is well known from classical dynamical critical
phenomena \cite{HohenbergHalperin77} that integrating out
soft modes leads to singularities in the resulting effective
theory.

Specifically, it was found \cite{BKV97} that the static spin susceptibility
$\chi^{(2)}({\bf r})$ behaves like $r^{-(2d-1)}$ for large distances
$r$. The leading long-wavelength dependence therefore has the form
\begin{eqnarray}
  \chi^{(2)}({\bf q}) / \chi^{(2)}(0) &=&
  1 + c_d (|{\bf q}|/2 k_F)^{d-1}  + O(|{\bf q}|^{2})
  \qquad(d<3) \label{eq:chi_d}
\end{eqnarray}
while in $d=3$ the non-analyticity takes the form
$c_3 (|{\bf q}|/2 k_F)^{2} \ln (2k_F/|{\bf q}|)$.
Here $k_F$ is the Fermi momentum and $c_d$ and $c_3$ are dimensionless
constants.
Note that these singularities only exist at zero temperature and
in zero magnetic field since both a finite temperature and a magnetic field
give the particle-hole excitations a mass.

Using (\ref{eq:chi_d}),
and with $\int_q = \sum_{\bf q} T\sum_{i\Omega}$,
the Gaussian part of $\Phi$ can be written,
\begin{equation}
\Phi^{(2)}[{\bf M}] = \int_q {\bf M}(q)\bigl[t_0
                       + c_{d}\vert{\bf q}\vert^{d-1} + c_2{\bf q}^2
                + c_\Omega \vert\Omega\vert/\vert{\bf q}\vert\bigr]\,
                                                 {\bf M}(-q)\quad.
\label{eq:9}
\end{equation}
Here $t_0 = 1 - \Gamma_t\chi^{(2)}({\bf q}\rightarrow 0,\omega_n = 0)$
is the bare distance from the critical point, and $c_\Omega$
is another constant.
Physically, the non-analytic term in the Gaussian part of $\Phi$
represents a long-range interaction of the spin
fluctuations which is self-generated by the electronic system.
For the same physical reasons for which the non-analyticity occurs in
$\chi^{(2)}$, the higher coefficients $\chi^{(n)}$ ($n>2$) in (\ref{eq:Hertz-LGW})
in general diverge for zero frequencies and wave numbers.
Consequently, the free energy functional (\ref{eq:Hertz-LGW})
is mathematically ill-defined. However, it will nonetheless be
possible to extract a considerable amount of information.

The sign of the non-analyticity in the Gaussian term merits some attention
since it will be responsible for the qualitative features of the ferromagnetic
quantum phase transition. Perturbation theory to second order in $\Gamma_t$
yields $c_{d}<0$ \cite{BKV97}. This is the generic case, and it
is consistent with the well-known
notion that correlation effects in general
decrease the effective Stoner coupling \cite{White}. However, Ref.\
\cite{BKV97} has given some possible mechanisms for $c_{d}$ to be
positive at least in some materials.

\subsection{Phase transition scenarios}

Depending on the sign of the non-analyticity in the
Gaussian term (\ref{eq:9}) of the free energy functional
there will be different scenarios for the ferromagnetic quantum
phase transition \cite{VBKN99}.

We first discuss the generic case of $c_{d}<0$. Here the free energy reduces
with increasing $q$ from zero which implies that a continuous transition to a
ferromagnetic state is impossible at zero temperature. Two possible scenarios
for the phase transition arise for $c_{d}<0$. The first scenario is based on
the observation that a finite thermodynamic magnetization $m=\langle |{\bf
M}(x)|\rangle$, which acts similarly to a magnetic field, cuts off
the singularities in the coefficients of the order parameter field theory.
Therefore, the non-analyticity in $\chi^{(2)}$ leads to an analogous
non-analyticity in the magnetic equation of state, which takes the form
\begin{eqnarray}
 t m - v_d m^d + u m^3 &=& H \qquad (d<3)~,
 \label{eq:equofstate1}\\
 t m - v_3 m^3 \ln(1/m) + u m^3 &=& H \qquad (d=3)~,
\label{eq:equofstate2}
\end{eqnarray}
where $t$ tunes the transition and $u, v_d$ and $v_3$ are positive constants.
$H$ denotes the external magnetic field.
This equation of state describes a first-order phase
transition since the next-to-leading term for small $m$ has a
negative sign.
This scenario was investigated in some detail in Ref.\ \cite{BKV99a}.
Since the non-analyticities
in $\chi^{(2)}$ and the equation of state are cut off by
a finite temperature, the transition will be of first order
at very low $T$ but turn second order at higher temperatures.
The two regimes are separated by a tricritical point. This is exactly
the behavior found experimentally in MnSi \cite{PML95,PMJL97}.

The second possible scenario for the quantum phase transition
arising if
$c_{d}<0$ is that the ground state of the system will not be ferromagnetic
but instead a spin-density wave at finite {\bf q}. This scenario has not been
studied in much detail so far, but work is in progress. It is tempting to
interpret the spiral ordering in MnSi as a signature of this finite-$q$
instability. This is, however, not very likely since a finite-$q$ instability
caused by the long-range interaction will be strongly temperature dependent
due to the temperature cutoff of the singularities. As mentioned above,
experimentally the ordering wave vector is essentially temperature
independent. Further work will be necessary to decide which of the two
possible scenarios, viz. a first-order ferromagnetic transition or a
continuous transition to modulated magnetic order, is realized under what
conditions. Moreover, let us point out, that in $d=3$ the non-analyticity is
only a logarithmic correction and would hence manifest itself only as a phase
transition at exponentially small temperatures, and exponentially large length
scales. Thus, it may well be unobservable experimentally for some materials.

We now turn to the second case, $c_{d}>0$ which can happen, if one of the
conditions discussed in Ref.\ \cite{BKV97} is fulfilled. In this case the
self-generated long-range interaction is ferromagnetic. Consequently, the
ferromagnetic quantum phase transition will be a conventional second order
phase transition, which can be analyzed by standard renormalization group
methods. A tree level analysis shows that the Gaussian theory is sufficient
for dimensions $d>d_c^+=1$ since all higher order terms are irrelevant. We are
therefore able to obtain the critical behavior exactly, yet due to the
long-range interaction it is {\em not} mean field-like.
The results of this analysis \cite{VBNK97} can be summarized as
follows. At zero temperature the equation of state close
to the quantum critical point reads
\begin{eqnarray}
 t m + v_d m^d + u m^3 &=& H \qquad (d<3)~, \label{eq:1a}\\
 t m + v_3 m^3 \ln(1/m) + u m^3 &=& H \qquad (d=3)~,
\label{eq:1b}
\end{eqnarray}
Again, $u$ and $v$ are positive constants. Note the different sign
of the non-analytic terms compared to eqs. (\ref{eq:equofstate1},
\ref{eq:equofstate2}).
From the equation of state
one obtains the critical exponents $\beta$ and $\delta$ while
the
correlation length exponent
$\nu$, the order parameter susceptibility exponent $\eta$, and the
dynamical exponent $z$ can be directly read of the Gaussian part
of $\Phi$, eq. (\ref{eq:9}). We find
$\beta = \nu = 1/(d-1)$, $\eta = 3-d$, $\delta = z = d$ for $1<d<3$.
These exponents `lock into' mean-field values
$\beta=\nu=1/2$, $\eta=0$, $\delta=z=3$ for $d>3$.
In $d=3$, there are logarithmic
corrections to power-law scaling.

At finite temperature, we find homogeneity laws for $m$, and for the
magnetic susceptibility, $\chi_m$,
\begin{equation}
m(t,T,H) = b^{-\beta/\nu} m(tb^{1/\nu}, Tb^{\phi/\nu}, Hb^{\delta\beta/\nu})
                                                                    \quad,
\label{eq:3a}
\end{equation}
\begin{equation}
\chi_m(t,T,H) = b^{\gamma/\nu} \chi_m(tb^{1/\nu}, Tb^{\phi/\nu},
                           Hb^{\delta\beta/\nu})\quad,
\label{eq:3b}
\end{equation}
where $b$ is an arbitrary scale factor. The susceptibility
exponent $\gamma$ and the crossover exponent $\phi$
that describes the crossover
from the quantum to the classical Heisenberg fixed point (FP) are given by
$\gamma = \beta (\delta - 1) = 1 ,\quad \phi = \nu$
for all $d>1$. Notice that the temperature dependence of the
magnetization is {\em not}
given by the dynamical exponent. However, $z$ controls the temperature
dependence of the specific heat coefficient, $\gamma_V = c_V/T$, which
has a scale dimension of zero for all $d$, and logarithmic corrections
to scaling for all $d<3$
\begin{equation}
\gamma_V(t,T,H) = \Theta(3-d)\,\ln b +
           \gamma_V(tb^{1/\nu}, Tb^z, Hb^{\delta\beta/\nu})\quad .
\label{eq:5}
\end{equation}

The singularities in the spin density correlation functions
do not only influence the properties of the quantum phase transition
but also those of the ferromagnetic phase. An example is
the dispersion relation of the ferromagnetic spin waves
\cite{BKMV98}.
Since the non-analyticities are cut off by a finite
magnetization it turns out that the dispersion relation
remains $\omega \propto q^2$ but the prefactor picks up
a non-trivial magnetization dependence (different from being
proportional to $m$ as in Stoner theory). For small
magnetization $m$ we find
\begin{eqnarray}
\omega &\propto& m^{d-2}\, q^2 \qquad\qquad\qquad (d<3)
\label{eq:spinwave-d}\\
\omega &\propto& m\ln(1/m)\, q^2 \qquad\qquad (d=3)~.
\label{eq:spinwave-3}
\end{eqnarray}
Until now, the corrections to mean-field theory predicted by
(\ref{eq:spinwave-d}) and (\ref{eq:spinwave-3}) have not
been observed experimentally.

\subsection{Influence of disorder}
\label{subsec:IFM_dis}

\begin{figure}[t]
  \epsfxsize=\textwidth
  \centerline{\epsffile{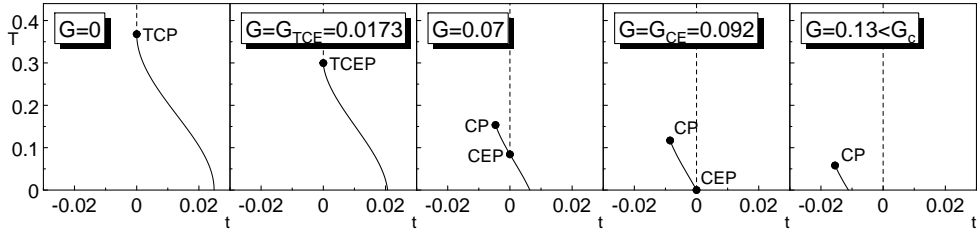}}
  \caption{Phase diagrams
  of disordered itinerant ferromagnets in the $T$-$t$-plane showing first
  order (solid lines) and second order (dashed lines)
  transitions. $G$ is a dimensionless
  measure of the disorder, and CP, CEP, TCP, and TCEP denote critical points,
  critical endpoints, tricritical points and tricritical endpoints,
  respectively. The first panel corresponds to the experimental results
  on MnSi \protect\cite{PML95,PMJL97}. See Ref.
  {\protect\cite{BKV99a}} for more details.}
  \label{fig:first}
\end{figure}

In this subsection we briefly discuss the influence of
quenched non-magnetic disorder on the ferromagnetic quantum
phase transition. An approach similar to that of the clean case has also
been developed for the dirty case \cite{KirkpatrickBelitz96}, and
the resulting effective theory
is very similar. Again, the magnetization couples to additional
soft modes (here with diffusive dynamics) which leads to an
effective long-range interaction. The singularities are even
stronger than in the clean case, but they have the opposite sign
so that the self-generated long-range interaction is generically ferromagnetic.
Thus in the presence of disorder there will be a
competition between the ballistic and diffusive singularities,
and the temperature which cuts off both.
For weak disorder the first-order transition will survive,
while larger disorder leads to a continuous transition.
As shown in Ref. \cite{BKV99a},
the phase diagram becomes very rich, showing
several multicritical points and even
regions with metamagnetic behavior (see Fig. \ref{fig:first}).
The properties of the continuous quantum phase transition
occurring for stronger disorder can again be analyzed by
standard renormalization group methods.
It turns out that as in the clean case the Gaussian
theory is sufficient since all higher order terms
are irrelevant. The resulting critical exponents are
$\gamma = 1$ for all $d>2$ and
$\nu = 1/(d-2)$, $\eta = 4-d$, $z = d$
for $2<d<4$. These exponents lock into their
mean field values  $\nu=1/2$, $\eta=0$, and $z=4$ for $d>4$.
In addition to $d=4$, $d=6$ also plays the role of an upper
critical dimension, and one has
$\beta = 2/(d-2)$, $\delta = d/2$
for $2<d<6$, while $\beta=1/2$, $\delta=3$ for $d>6$.

\section{Influence of rare regions on
         magnetic quantum phase transitions}
\label{sec:RR}

\subsection{Disorder, rare regions, and the Griffiths region}
\label{subsec:griffiths}

The influence of static or quenched disorder on the critical properties
of a system near a continuous phase transition is a very interesting
problem in statistical mechanics. While it was initially suspected that
quenched disorder always destroys any critical point \cite{Grinstein85},
this was soon found to not necessarily be the case. Harris \cite{Harris74,CCFS86}
found a convenient criterion for the stability of a given critical
behavior with respect to quenched disorder: If the correlation length
exponent $\nu$ obeys the inequality $\nu\geq 2/d$, with $d$ the spatial
dimensionality of the system, then the critical behavior is unaffected
by the disorder. In the opposite case, $\nu < 2/d$, the disorder modifies
the critical behavior \cite{controversy}.
This modification may either (i) lead to a new critical
point that has a correlation length exponent $\nu \geq 2/d$ and is thus
stable, or (ii) lead to an unconventional critical point where the
usual classification in terms of power-law critical exponents looses
its meaning, or (iii) lead to the destruction of a sharp phase transition.
The first possibility is realized in the conventional theory of random-$T_c$
classical ferromagnets \cite{Grinstein85}, and the second one is probably
realized in classical ferromagnets in a random
field \cite{Villain85,Fisher86,NKJ91}. The third one has occasionally been attributed to
the exactly solved McCoy-Wu model \cite{McCoyWu68,McCoy69,ShankarMurthy87}.
This is misleading,
however, as has recently been emphasized in Ref.\ \cite{Fisher95};
there is a sharp, albeit unorthodox, transition in that
model, and it thus belongs to category (ii).

Independent of the question of if and how the critical behavior is affected,
disorder leads to very interesting phenomena as a phase transition is
approached. Disorder in general decreases the critical temperature $T_c$ from
its clean value $T_c^0$. In the temperature region $T_c<T<T_c^0$ the system
does not display global order, but in an infinite system one will find
arbitrarily large regions that are devoid of impurities, and hence show
local order, with a small but non-zero probability that usually decreases
exponentially with the size of the region. These static disorder fluctuations
are known as `rare regions', and the order parameter fluctuations induced
by them as `local moments' or `instantons'. Since they are weakly coupled,
and flipping them requires to change the order parameter in a whole
region, the local moments have very slow dynamics.
Griffiths \cite{Griffiths69} was the first to show that they lead to a
non-analytic free energy everywhere in the region $T_c<T<T_c^0$, which
is known as the Griffiths phase, or, more appropriately, the Griffiths
region. In generic classical systems this is a weak effect, since the
singularity in the free energy is only an essential one. An important
exception is the McCoy-Wu model \cite{McCoyWu68}, which is a $2D$
Ising model with bonds that are random along one direction, but identical
along the second direction. The resulting infinite-range correlation of
the disorder in one direction leads to very strong effects.
As the temperature is lowered through the Griffiths
region, the local moments cause the divergence of an increasing number of
higher order susceptibilities,
$\partial^n m/\partial B^n$ ($n\geq 2$), starting with large $n$.
Even the average susceptibility proper, $\chi^{(2)}=\partial m/\partial B$,
diverges at a
temperature $T_{\chi} > T_c$, although the average order parameter does
not become non-zero until the temperature
reaches $T_c$. This is caused by rare fluctuations in the susceptibility
distribution, which dominate the average susceptibility and make it very
different from the typical or most probable one.

Surprisingly little is known about the influence of the Griffiths region
and related phenomena on the critical behavior. Recent work \cite{DHSS95}
on a random-$T_c$ classical Ising model has suggested that it can be profound,
even in this simple model where the conventional theory predicts standard
power-law critical behavior, albeit with critical exponents that are
different from the clean case. The authors of
Ref.\ \cite{DHSS95} have shown that the conventional theory is
unstable with respect to perturbations that break the replica symmetry.
By approximately taking into account the rare regions, which are
neglected in the conventional theory, they found a new term in the
action that actually induces such perturbations. In some systems
replica symmetry breaking is believed to be associated with activated,
i.e. non-power law, critical behavior. Reference \cite{DHSS95}
thus raised the interesting possibility that, as a result of rare-region
effects, the random-$T_c$ classical Ising model shows activated critical
behavior, as is believed to be the case for the random-field classical
Ising model \cite{Villain85,Fisher86,NKJ91},
although in the case of the random-$T_c$
model no final conclusion about the fate of the transition could be
reached.

Griffiths regions also occur in the case of quantum phase transitions
(for an experimental example see Ref.\ \cite{ACDDFGMMCC98}). Their consequences for
the critical behavior are even less well investigated than in the
classical case, with the remarkable exception of certain
$1D$ systems. Fisher \cite{Fisher95} has
investigated quantum Ising spin chains in a transverse random field.
These systems are closely related to the classical McCoy-Wu model, with
time in the quantum case playing the role of the `ordered direction' in
the latter. He has found activated critical
behavior due to rare regions. This has been confirmed by numerical
simulations \cite{KiskerYoung98}.
Other recent simulations \cite{PYRK98} suggest that this
type of behavior may not be restricted to $1D$ systems, raising
the possibility that exotic critical behavior dominated by rare regions
may be generic in quenched disordered quantum systems, independent of the
dimensionality and possibly also of the type of disorder.

\subsection{Itinerant quantum antiferromagnets}
\label{subsec:IQAFM}

Within the conventional theory \cite{Grinstein85} of
critical behavior in systems with quenched disorder the first
step consists of averaging over the disorder, usually via the replica
trick \cite{EdwardsAnderson75}. The resulting effective theory is then
analyzed perturbatively. However, the rare regions are a non-perturbative
effect since the probability for their occurrence is exponentially small
in the disorder strength. Therefore, rare regions are neglected within
the conventional theory.

Narayanan, Vojta, Belitz, and Kirkpatrick \cite{NVBK99} have developed
a generalization of the conventional theory of quantum phase transitions
in the presence of quenched disorder. This theory, which is similar to
that of Ref.\ \cite{DHSS95} for classical transitions, includes the effects
of the rare regions. The basic idea is {\em not} to average over the disorder
at the beginning but to work with a particular disorder configuration
until the rare regions are identified. Only after their effects have been
incorporated into the theory, the disorder average is carried out.

In the following we illustrate this theory taking the itinerant
quantum antiferromagnet as the primary example. The starting point
is the order parameter field theory for the itinerant quantum
antiferromagnet derived by Hertz \cite{Hertz76}. The
Landau-Ginzburg-Wilson free energy functional reads
\begin{equation}
S = \int dx\,dy\ {\boldphi}(x)\,\Gamma_0(x,y)\,{\boldphi}(y)
    + u\int dx\ \left({\boldphi}(x)\cdot{\boldphi}(x)\right)^2\,,
\label{eq:Hertz-AFM}
\end{equation}
where $\boldphi$ is the staggered magnetization. $\Gamma(x,y)$ is the bare
two-point vertex function, whose Fourier transform is
\begin{equation}
\Gamma_0({\bf q},\Omega_n) = (t_0 + {\bf q}^2 + \vert\omega_n\vert)\quad.
\end{equation}
Disorder is introduced by making the distance $t$
from the critical point a random function of position,
$t({\bf x}) = t_0 + \delta t({\bf x})$, where $\delta t({\bf x})$ obeys a
Gaussian distribution with zero mean and variance $\Delta$.

Instead of averaging over the disorder we now determine saddle point
solutions of the unaveraged Landau-Ginzburg-Wilson functional
(\ref{eq:Hertz-AFM}). Due to the disorder there will be spatial regions in
which the system wants to order ($t({\bf x})<0$) even if it is globally in its
disordered phase ($t_0>0$). These rare regions or islands will support
locally nonzero saddle-point solutions. Outside of the islands,
the solution is exponentially small. Thus, the islands are
effectively decoupled. For a system with
$N$ islands, and in the case of Ising symmetry, there will be $2^N$
almost degenerate saddle-point solutions that can be constructed by
considering all possible distributions of the sign of the order parameter
on the islands. For a continuous order parameter symmetry
there is a whole manifold of almost degenerate saddle points.
This complicated structure of the free energy landscape is responsible
for the failure of the conventional theory as is known from the
random field Ising model \cite{Villain85,Fisher86,NKJ91}.

Now, the crucial point is that for a complete theory one has to take
into account fluctuations around {\em all} of these saddle points.
As was shown in Ref.\ \cite{NVBK99} the saddle point configurations
act as an additional source of disorder in the system. Since the saddle
points are time-independent this disorder is static, but it is self-generated
and thus in equilibrium with the rest of the system. Therefore, taking into
account fluctuations around all saddle points leads to the appearance of
static annealed disorder in addition to the underlying quenched disorder.
(Some general aspects concerning annealed disorder and quantum phase transitions
are discussed in Ref.\ \cite{NVBK99a}.)
At this point in the calculation
the average of the quenched disorder is carried out by
means of the replica trick. The resulting effective theory for
the fluctuations $\boldvarphi$ takes the form
\begin{eqnarray}
{S}_{\rm eff}&=&~\sum_{\alpha} \int dx\,dy\ \boldvarphi^{\alpha}(x)\,
                   \Gamma_{0}(x,y)\,\boldvarphi^{\alpha}(y)
+ u\sum_{\alpha}\int dx\ \left(\boldvarphi^{\alpha}(x)\cdot
                                    \boldvarphi^{\alpha}(x)\right)^2
\nonumber\\
&-& \Delta \sum_{\alpha,\beta} \int dx\,dy\,
  \delta({\bf x}-{\bf y})\,
  \left(\boldvarphi^{\alpha}(x)\right)^2\,\left(\boldvarphi^{\beta}(y)\right)^2
 \nonumber\\
&-& T {\bar w} \sum_{\alpha} \int dx\,dy\,
  \delta({\bf x}-{\bf y})\,
  \left(\boldvarphi^{\alpha}(x)\right)^2\,\left(\boldvarphi^{\alpha}(y)\right)^2~.
\label{eq:AFM_Seff}
\end{eqnarray}
Here the first line represents the clean antiferromagnet, the second is the
conventional quenched disorder term and the last line contains the static
annealed disorder which is due to the rare regions. The temperature
factor in front of the annealed disorder term originates from the Boltzmann
factor
for the saddle point free energy. The parameter $\bar w$ contains the
probability for finding rare regions and the strength of the local order
on the islands.
Since $\bar w$ is non-perturbative in the disorder strength, the theory
contains effects beyond the conventional perturbative approach.

This effective translationally invariant theory can now be analyzed by
standard renormalization group methods. It turns out \cite{NVBK99}
that the new term in the Landau-Ginzburg-Wilson functional (\ref{eq:AFM_Seff})
destabilizes the critical fixed point found within the conventional
theory \cite{KirkpatrickBelitz96b}. No new fixed point is found (at one loop
order of the perturbation theory). Instead, the system displays runaway flow
to large disorder values in the entire physical parameter space.
Within the renormalization group approach it is not possible to determine the
ultimate fate of the transition. The runaway flow
can be interpreted either as a complete destruction
of the antiferromagnetic long-range order in favor of a random singlet phase
\cite{BhattLee82,BhattFisher92} or the existence of a non-conventional critical
point (e.g., with activated scaling).

While rare regions destroy the conventional critical point in
itinerant quantum antiferromagnets they do not influence the quantum
phase transition of itinerant {\em ferromagnets} \cite{NVBK99}.
The reason is the effective long-ranged interaction between the order
parameter fluctuations discussed in Sec.\ \ref{sec:IQM}. It suppresses
all fluctuations including those generated by the rare regions. Therefore
the conventional critical behavior discussed in Subsec.\ \ref{subsec:IFM_dis}
will not be changed by the rare regions.

\section{Metal-insulator transitions of disordered interacting electrons}
\label{sec:MIT}

\subsection{Localization and interactions}

Metal-insulator transitions are a particularly fascinating and only
incompletely understood class of quantum phase transitions.
Conceptually, one distinguishes between
Anderson transitions in models of noninteracting electrons, and Mott-Hubbard
transitions of clean, interacting electrons. At the former, the
electronic charge diffusivity $D$ is driven
to zero by quenched, or frozen-in, disorder, while the thermodynamic
properties do not show critical behavior. At the latter,
the thermodynamic density susceptibility $\partial n/\partial\mu$
vanishes due to electron-electron interaction effects. In either case, the
conductivity $\sigma = (\partial n/\partial\mu)D$ vanishes at the
metal-insulator transition.

The investigation of the disorder-driven metal-insulator transition
has a long history. Anderson \cite{Anderson58} was the first to realize
that introducing quenched disorder into a metallic system, e.g., by adding
impurity atoms, can change the nature of the electronic states from spatially
extended to localized.
This localization transition of disordered {\em non-interacting} electrons,
the Anderson transition, is comparatively well
understood (for a review see, e.g., Ref.\ \cite{KramerMacKinnon93}).
The scaling theory of localization \cite{AALR79} predicts that
in the absence of spin-orbit coupling or magnetic fields
all states are localized in one and two spatial dimensions for arbitrarily weak
disorder. Thus, no true metallic phase exists in these dimensions.
In contrast, in three dimensions there is a phase transition from
extended states for weak disorder to localized states for strong disorder.
These results of the scaling theory are in agreement with large-scale
computer simulations of non-interacting disordered electrons.

However, in reality electrons do interact via the Coulomb potential,
and the question is, how this changes the above conclusions.
The conventional approach to the problem of disordered {\em interacting}
electrons is based on a perturbative treatment of both disorder and
interactions (for reviews see, e.g., Refs.\
\cite{AltshulerAronov85,LeeRamakrishnan85}).
It leads to a scaling theory and a related
field-theoretic formulation of the problem \cite{Finkelstein83}, which was
later investigated in great detail within the framework of the renormalization
group (for a review see Ref.\ \cite{BelitzKirkpatrick94}).
One of the main results is that in the absence of external symmetry-breaking
(spin-orbit coupling or magnetic impurities, or a magnetic field)
a phase transition between a {\em normal} metal
and an insulator only exists in dimensions larger than two, as was the case for
non-interacting electrons. In two dimensions the results of this approach
are inconclusive since the renormalization group displays runaway flow to
zero disorder but infinite interactions. Furthermore, it has not been
investigated so far, whether effects of rare regions analogous to those
discussed in Sec.\ \ref{sec:RR} for magnetic transitions
would change the above conclusions about the metal-insulator transition.

\begin{figure}[t]
  \epsfxsize=7cm
  \centerline{\epsffile{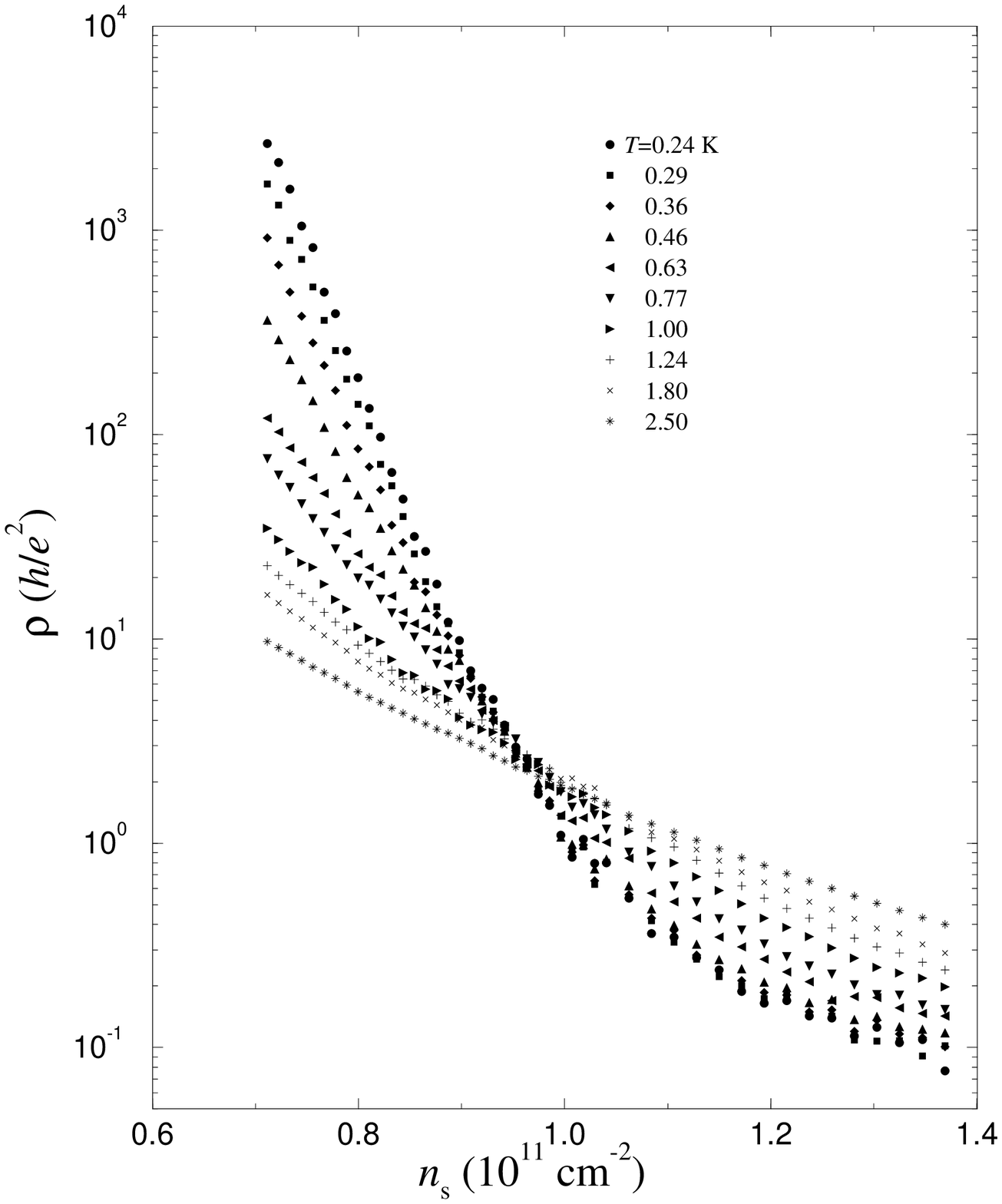}\hspace{-0.4cm}\epsfxsize=7cm\epsffile{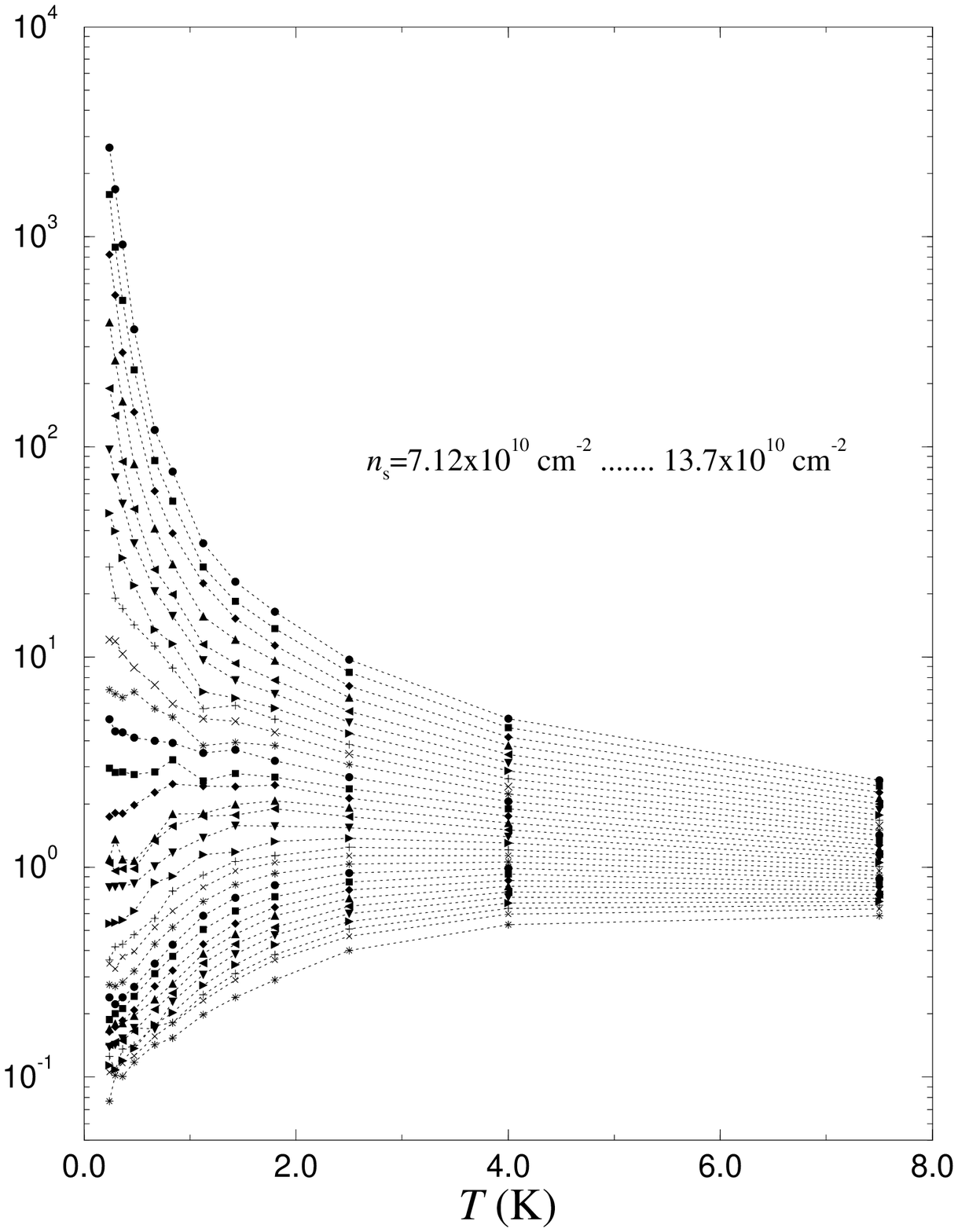}}
  \caption{Resistivity of the 2D electron gas a Si-MOSFET as function
          of carrier concentration and doping. The data clearly indicate
          the existence of a metal-insulator transition. (from Ref.\
          \protect\cite{KMBFPD95})}
  \label{fig:2DMIT}
\end{figure}

Experimental work on the disorder-driven metal-insulator transition
(mostly on doped semiconductors) carried out before 1994 essentially
confirmed the existence of a transition in three dimensions while no
transition was found in two-dimensional systems.
Therefore it came as a surprise when experiments
on Si-MOSFETs \cite{KKFPD94,KMBFPD95} revealed indications of a
true metal-insulator transition
in two dimensions (see Fig.\ \ref{fig:2DMIT}).
While these results were first viewed with considerable skepticism
they were soon confirmed \cite{PFW97} and later also found in various
other materials \cite{2DMITother}.\footnote{Note, however,
that very recent experimental data
\cite{SHPLRR99} on 2d GaAs
hole systems indicate that the seeming metallic phase
is a finite temperature phenomenon. For sufficiently low temperatures
the old results of scaling theory remain valid, and thus there may be
no true metallic phase in two dimensions at least in this material.}
It soon became clear that the main difference between the
new experiments and those carried out earlier was that the electron
(or hole) density is very low.
Therefore, the Coulomb interaction is particularly strong
compared to the Fermi energy.
For example, in the Si-MOSFETs the typical electron density is
$10^{11}{\rm cm}^{-2}$ leading to a typical Coulomb energy of about 10 meV
while the Fermi energy is only about 0.5 meV.
Therefore interaction effects are a likely reason for this new
metal-insulator transition in two dimensions.
A complete understanding
has, however, not yet been obtained. Different explanations
have been suggested
based on the perturbative renormalization group \cite{CDL98,SiVarma98},
non-perturbative effects \cite{CYA98,CKNV98}, or the transition
actually being a superconductor-insulator transition rather than a
metal-insulator transition \cite{SIT}. In addition to these interaction
based explanations a number of more conventional suggestions have been
made, among them the presence of temperature-dependent disorder
as provided by the filling and emptying of charge traps
\cite{AltshulerMaslov99} and temperature-dependent screening
\cite{KlapwijkDasSarma98}.

This is not the place to discuss all these developments in detail
or even to review the vast field of metal-insulator transitions.
Instead, we will concentrate on a few aspects of the metal-insulator
transition in the presence of both disorder and interactions.

\subsection{Rare regions, local moments, and annealed disorder,
            a new mechanism for the metal-insulator transition}

In this subsection we discuss how rare regions analogous to those
studied in Sec.\ \ref{sec:RR} influence the
metal-insulator transition. Let us consider an electron system in the
presence of both interactions and (nonmagnetic) quenched disorder. Due to the disorder
there will be rare spatial regions where the exchange interaction
is greatly enhanced. In these regions the system will display local
magnetic order. Physically, these regions correspond to local magnetic
moments. There is much experimental evidence for local moments
\cite{LM_ex}, and their formation has been studied theoretically
\cite{LM_th}.

Belitz, Kirkpatrick and Vojta \cite{BKV99} have developed an approach which
includes the effects of the local moments into a transport theory, starting
from a field-theoretical description of disordered interacting electrons.
As in Sec.\ \ref{sec:RR} the general idea is to avoid the disorder average
at the beginning of the calculation but to work with a fixed disorder
configuration. This leads to the appearance of spatially inhomogeneous
saddle points of the field theory. In particular, there will be saddle points
which have non-zero magnetization in some rare spatial regions.
Analogous to Sec.\ \ref{sec:RR} summing over the manifold of degenerate
saddle points leads to the appearance of {\em annealed} magnetic disorder
in addition to the underlying (nonmagnetic) quenched disorder. Let us emphasize
that this annealed magnetic disorder is generically self-generated by the
system.

In Ref.\ \cite{BKV99} the annealed magnetic disorder was then incorporated
into the sigma-model description of the metal-insulator transition. To simplify
the problem, a model of non-interacting electrons with annealed magnetic
disorder was considered. This corresponds to neglecting all interaction
effects beyond the formation of local moments. The resulting non-linear
sigma  model can be analyzed using the standard renormalization group methods
\cite{BelitzKirkpatrick94}. It turns out that the annealed magnetic disorder
leads to a new mechanism and a new universality class for the metal-insulator
transition which is different from the conventional localization transition.
Note that the effects of {\em annealed} magnetic disorder are also very different
from the case of quenched magnetic impurities.
For the simplified model and neglecting the Cooper channel we find
that the diffusion coefficient is not renormalized at one loop order
while the thermodynamic density susceptibility
$\partial n / \partial \mu$ is driven to zero. Thus, the transition resembles
a Mott-Hubbard transition rather than an Anderson transition.

The lower critical dimension for this new transition is two, as it is for
the conventional localization transition. In two dimensions the system is
insulating for any disorder. Therefore, the local moments
alone do not provide an explanation for the metal-insulator transition
in the two-dimensional electron system in Si-MOSFETs and other materials
discussed in the last subsection. Clearly, it would be interesting to study
generalizations of the model studied in Ref.\ \cite{BKV99} which include
the Cooper channel and interactions beyond the formation of local moments.

\subsection{Numerical simulation of disordered interacting electrons}

The remaining part of Section \ref{sec:MIT} is devoted to the numerical work
on interacting electrons in the presence of quenched
disorder.  The model investigated is the quantum Coulomb glass
model \cite{Efros95,TPE96,ESV97},
a generalization of the classical Coulomb glass model
\cite{Pollak70,EfrosShklovskii75} which was used to study disordered insulators.
The quantum Coulomb glass is defined on a hypercubic lattice of $L^d$
sites occupied by $N=K\, L^d $ spinless electrons ($0\!<\!K\!<\!1$).
To ensure charge neutrality each lattice site carries
a compensating positive charge of  $Ke$. The Hamiltonian
is given by
\begin{equation}
H =  -t  \sum_{\langle ij\rangle} (c_i^\dagger c_j + c_j^\dagger c_i) +
       \sum_i \varphi_i  n_i + \frac{1}{2}\sum_{i\not=j}(n_i-K)(n_j-K)U_{ij}
\label{eq:Hamiltonian}
\end{equation}
where $c_i^\dagger$ and $c_i$ are the electron creation and annihilation operators
at site $i$, respectively,  and $\langle ij \rangle$ denotes all pairs of nearest
neighbor sites.
$t$ gives the strength of the hopping term and $n_i$ is the occupation number of site $i$.
For a correct description of the insulating phase the Coulomb
interaction between the electrons is kept long-ranged,
$U_{ij} = U/r_{ij}$, since screening breaks down in the insulator
(the distance $r_{ij}$ is measured in units of the lattice constant).
The random potential values $\varphi_i$ are chosen
independently from a box distribution of width $2 W$ and zero average.
Two important limiting cases of the quantum Coulomb glass are the Anderson model of
localization (for $U_{ij}=0$) and the classical Coulomb glass (for $t=0$).

For two reasons the numerical simulation of disordered quantum many-particle
systems is one of the most complicated
problems in computational condensed matter physics. First, the dimension of the
Hilbert space to be considered grows exponentially with the system size. Second, the presence
of quenched disorder requires the simulation of many samples with different disorder configurations
in order to
obtain averages or distribution functions of physical quantities. In the case of
disordered interacting electrons the problem is even more challenging due to
the long-range character of the Coulomb interaction
which has to be retained, at least for a correct description of  the insulating
phase. Here we discuss the results of two different numerical methods to
tackle the problem. First, the Coulomb interaction is decoupled by
means of
a Hartree-Fock approximation and numerically solved the remaining
self-consistent disordered single-particle problem. This method
permits comparatively large system sizes of more than $10^3$ sites.
The results of this approach are
summarized in Sec.\ \ref{subsec:HF} together with those of exact diagonalization
studies we performed to check the quality of the Hartree-Fock approximation.
Since the Hartree-Fock method turned out to be a rather poor approximation for
the calculation of transport properties an
efficient method to calculate the low-energy properties of
disordered quantum many-particle systems with high accuracy has been developed.
This method, the Hartree-Fock based diagonalization, and the
results we have obtained this way are summarized in Sec.\ \ref{subsec:HFD}.

\subsection{Hartree-Fock approximation}
\label{subsec:HF}

The Hartree-Fock approximation consists in decoupling the Coulomb interaction
by replacing operators by their expectation values:
\begin{eqnarray}
H_{\rm HF} =&-&t  \sum_{\langle ij\rangle} (c_i^\dagger c_j + c_j^\dagger c_i)
+  \sum_i  (\varphi_i -\mu) n_i \nonumber \\
 &+& \sum_{i\not=j} n_i ~ U_{ij} \langle n_j -K \rangle
 - \sum_{i,j} c_i^\dagger c_j ~ U_{ij} \langle c_j^\dagger c_i \rangle,
\label{eq:HF}
\end{eqnarray}
where the first two terms contain the single-particle part of the Hamiltonian, the
third is the Hartree energy and the fourth term contains the exchange interaction.
$\langle \ldots \rangle$ represents the expectation value with respect to
the Hartree-Fock ground state which has to be determined self-consistently.
In this way the many-particle problem
is reduced to a self-consistent disordered single-particle problem which we
solve by means of numerically exact diagonalization.

This method was applied to the three-dimensional quantum Coulomb glass model
\cite{ESV97}. It was found that the
interaction induces a depletion of the single-particle density of states
in the vicinity of the Fermi energy. For small hopping strength $t$
the depletion takes the form of a Coulomb gap
\cite{EfrosShklovskii75,EfrosShklovskii85} known from the classical ($t=0$)
limit. With increasing hopping strength there is a crossover from the
nearly parabolic Coulomb gap to a square root singularity characteristic
of the Coulomb anomaly \cite{AltshulerAronov79} in the metallic limit.
The depletion of the density of states at the Fermi energy has drastic
consequences for the localization properties of the electronic states.
Since the degree of
localization is essentially determined by the ratio between the hopping
amplitude and the level spacing, a reduced density of states directly leads
to stronger localization. Specifically, we calculated the inverse
participation number
\begin{equation}
P_\nu^{-1} = \sum_j |\langle j|\nu \rangle|^4
\label{eq:invPN}
\end{equation}
of a single-particle state $|\nu\rangle$ and compared the cases of non-interacting and
interacting electrons. In the presence of interactions we found a
pronounced maximum at the Fermi energy with values above that of
non-interacting electrons. Thus, within the Hartree-Fock approximation
electron-electron interactions lead to enhanced localization.

In order to precisely
determine how the location of the metal-insulator transition changes as a result
of this effect, we used
the fact that the spectral statistics on the insulating side
is equivalent to that of a Poisson ensemble of random matrices while
the spectral statistics on the metallic side is that of a Gaussian
orthogonal ensemble of random matrices \cite{SSSLS93,HofstetterSchreiber93}.
By analyzing the
statistics of the Hartree-Fock levels as a function of the hopping
amplitude $t$ we could thus determine the location of the
metal-insulator transition in the phase diagram.
In all cases the transition to a metal requires a larger hopping strength
for interacting electrons than for non-interacting.

Since the Hartree-Fock method is an uncontrolled approximation
it is highly desirable to compare these results to those of
exact many-particle calculations. Now the question arises,
what quantities are particular suitable for such a comparison.
In principle, one should compare the values of physical
observables like the conductivity
which is given by the Kubo-Greenwood formula \cite{Kubo57,Greenwood58}
\begin{equation}
 {\rm Re} ~ G^{xx}(\omega) = \frac {2 \pi^2}  {\omega} \sum_{\nu} |\langle 0 | j^x|\nu \rangle |^2
     \delta(\omega+E_0-E_{\nu})
\label{eq:kubo}
\end{equation}
where $\omega$ denotes the frequency.
$j^x$ is the $x$ component of the current operator and $|\nu\rangle$ denotes the eigenstates
of the Hamiltonian.
However, the Kubo-Greenwood conductivity is a complicated
quantity from a numerical point of view since it involves
a nontrivial extrapolation to zero frequency. Therefore,
it is desirable to apply the simpler localization criteria
known from non-interacting systems. This leads, however, to
an additional complication.
The usual criteria which are based on the
(single-particle) participation number
or the statistics of (single-particle) energy levels
are not even defined for many-particle states. While this does
not create any problems at the Hartree-Fock level
(the Hartree-Fock states are effective single-particle states)
it implies that the criteria cannot directly be applied to
true many-particle states. There have been several suggestions
(for a discussion see Ref.\ \cite{TES98}) how to generalize
the participation number to a many-particle state. It turns out
that a unique procedure does not exist, and different generalizations
have different merits.
Since localization is connected with the probability of a particle to
return to its starting position, a promising
candidate to replace the inverse participation number is
\begin{equation}
R_p(\varepsilon) = \frac 1 {g(\varepsilon)} \frac 1 N \sum_j
     \lim_{\delta \to 0} \frac \delta \pi \, G_{jj}(\varepsilon + i \delta)
     \, G_{jj}(\varepsilon - i \delta) ~.
\end{equation}
Here $g(\varepsilon)$ is the single-particle density of states and
$G_{jj}$ is the diagonal element of the single-particle Greens function.
This quantity measures the probability of a single particle excitation
created at site $j$ to return to this site in infinite time.
However, in addition to the desired localization information
$R_p$ also contains information about the decay of
the quasi-particles. To disentangle localization and decay information
we note that the quasi-particle weight $z_0$ (which measures
the decay only) can be expressed as
\begin{equation}
z_0^2(\varepsilon) = \frac 1 {g(\varepsilon)} \frac 1 N \sum_{ij}
     \lim_{\delta \to 0} \frac \delta \pi \, G_{ij}(\varepsilon + i \delta)
     \, G_{ji}(\varepsilon - i \delta) ~.
\end{equation}
The desired localization information is thus contained in the normalized
return probability
$P^{-1}=R_p / z_0^2$ which is the natural generalization of the inverse
participation number (\ref{eq:invPN}) to interacting systems. For non-interacting electrons
it reduces to the inverse participation number proper.

Using these criteria the Hartree-Fock results were compared \cite{ESV98} to
those of numerically exact diagonalizations of the many-particle
Hamiltonian. Because the
exact calculations require the diagonalization of a matrix whose
dimension equals the size of the Hilbert space this comparison
was restricted to systems with not more than 16 sites.
While it was found that the
Hartree-Fock approximation yields reasonable results for static
quantities like the single-particle density of states, it does a very poor job
for localization properties and for time correlation functions
such as the conductivity. Thus, the Hartree-Fock results for the influence
of the electron-electron interaction on transport quantities
are highly unreliable. For this reason it was necessary to develop a new numerical
method, the Hartree-Fock based diagonalization which will be discussed
in the next subsection.

\subsection{Hartree-Fock based diagonalization}
\label{subsec:HFD}

Since numerically exact diagonalizations of the full many-particle
Hamiltonian are restricted to very small system
sizes and, as we have seen, the Hartree-Fock results for transport
properties are unreliable, a different method which gives exact results or
at least provides a controlled approximation for comparatively large systems
is highly desirable.
Such a method is the Hartree-Fock based diagonalization \cite{VES98}.
It is related to the
configuration interaction (CI) approach used in quantum chemistry (see, e.g.,
Ref.\ \cite{Fulde95}). The basic idea is to diagonalize the many-particle Hamiltonian
not in a real-space or momentum basis but rather in an energetically
truncated basis of Hartree-Fock states.
The Hartree-Fock states are comparatively close in character to the exact
eigenstates in the entire
parameter space. Therefore it is sufficient to keep only a very small
fraction of the Hilbert space (e.g., 2000 out of  $9\times 10^9$ states for 18 electrons on 36
lattice sites) to obtain low-energy  quantities with an accuracy comparable to
that of exact diagonalization.
A schematic of the Hartree-Fock based
diagonalization method is shown in Fig.\ \ref{fig:HFD}.
\begin{figure}
\begin{center}
\fboxrule1pt
\fboxsep5mm
\fbox{\parbox[c]{140mm}{\tt
\begin{tabbing}
\= do \= for each disorder configuration \hfill\\
\>   \>  solve  HF approximation\\
\>   \>  construct many-particle HF states\\
\>   \>  find lowest-in-energy HF states \\
\>   \>  transform Hamiltonian to basis of low-energy HF states\\
\>   \>  diagonalize Hamiltonian\\
\>   \>  transform observables to HF basis and calculate their values\\
\> enddo
\end{tabbing}
}}
\end{center}
\caption{Structure of the Hartree-Fock (HF) based diagonalization method.}
\label{fig:HFD}
\end{figure}
So far we have carried out test calculations for systems with up to
64 lattice sites (32 electrons) and production runs for up to
36 lattice sites (18 electrons).

\begin{figure}
  \epsfxsize=10cm
  \centerline{\epsffile{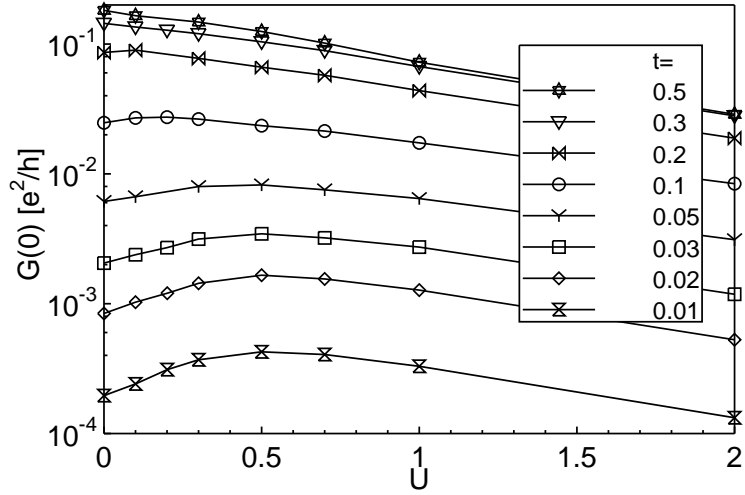}}
  \caption{d.c.\ conductance of a 2D system of $5\times 5$ sites. The disorder
    strength is $W=1$. The results represent logarithmic averages over 400
    disorder configurations. (from Ref.\ \protect\cite{VES98a})}
  \label{fig:gzer2d}
\end{figure}

As a first application of the Hartree-Fock based diagonalization method
the quantum Coulomb glass was studied in two dimensions \cite{VES98a}.
In order to determine
the influence of the Coulomb interaction on transport
the Kubo-Greenwood conductance (\ref{eq:kubo}) was calculated.
It turned out that the influence of
electron-electron interactions on the d.c.\ conductance is opposite
in the weakly and strongly disordered regimes.
The conductance of strongly disordered electrons is considerably enhanced
by a weak interactions. With increasing kinetic energy the relative
enhancement decreases as does the interaction range where the enhancement
occurs. The conductance of weakly disordered electrons is reduced even
by weak interactions.
In contrast, sufficiently strong interactions always reduce the conductance,
and the system approaches a Wigner crystal or Wigner glass state.
These results are summarized in Fig.\ \ref{fig:gzer2d}.

The qualitative difference we found between the weakly and strongly
disordered regimes can be explained in terms of two competing
effects of the interactions. The dominant effect in the case
of strong disorder is the destruction of phase coherence for
the {\em single-electron} motion.
It can be understood from the following simplistic argument:
Electron-electron collisions are inelastic and thus phase-breaking
from the point of view of a single electron. Therefore collisions destroy
the quantum interference necessary for Anderson
localization.\footnote{
Of course, this argument is an oversimplification: Taken at face value,
it appears to {\em always} predict diffusion and thus metallic behavior.
In order to obtain a more detailed prediction one has to analyze the
phase space for electron-electron collisions in the localized regime.
Moreover, what is neglected in the above argument is that the energy
and phase of the many-particle state are not changed by a collision.
}
It is clear that this mechanism to increase the conductance is
particularly effective if the localization length is very small to
begin with.
In the opposite case, i.e., for weak disorder, the
dominant effect of the interactions
is an increasing pinning of the electrons by the
repulsive forces which for strong interactions eventually leads
to the formation of a Wigner glass or crystal.
This simple arguments also explain, why the Hartree-Fock approximation
did not reproduce the enhanced conductance in the localized regime:
The Hartree-Fock potential is static and therefore it does not break
the phase of the electron -- in contrast to real electron-electron collisions.

Analogous investigations were performed for quantum Coulomb glasses in
three \cite{VojtaEpperlein98} and one \cite{SEV99} spatial
dimensions. It turned out that the qualitative disorder
and interaction dependencies of the conductance are identical to those
in two dimensions. In particular, it was always found that weak interactions
induce delocalization in the strongly disordered regime.
The degree of delocalization seems to be determined essentially by
the ratio between disorder strength $W$ and the hopping band width
$z t$ where $z$ is the number of nearest neighbor sites.
In addition to the results obtained by the HFD method, other numerical
simulations also show a delocalizing influence of electron-electron
interactions in one \cite{SJWP98} and two \cite{Efros95,TPE96,BWP99,DST99}
dimensions.
The numerical results for the influence of weak interactions on the
Kubo-Greenwood conductance can be compared to the results from the
perturbative renormalization group for spinless fermions
\cite{ApelRice82} which predicts
that repulsive interactions always reduce
the transport of disordered spinless fermions. Thus, the numerical
results agree with those from perturbation theory in the case of
weak disorder, while in the strongly disordered regime the perturbative
results are qualitatively incorrect.

In addition to the Kubo-Greenwood conductance
the above mentioned return probability of single-particle excitations
was also calculated.
The return probability at the Fermi
energy displays a similar behavior as the Kubo-Greenwood conductance:
For weak disorder the return probability increases with the interaction
strength. The delocalizing tendency in the strongly localized regime
is less pronounced, probably because the Coulomb gap in the single-particle
density of states counteracts the delocalization.

In summary, the results obtained so far clearly show that interactions
can have a delocalizing tendency in certain parameter regions. However,
it is not
clear whether this is sufficient to explain the metal-insulator transition
found in two dimensions. More detailed investigations are presently under way
to find the finite-size scaling behavior of the conductance. Moreover,
we are currently generalizing the simulations to include spin degrees
of freedom which have been shown to be important in the experiments
on Si-MOSFETs \cite{SKS97,PBPB97}.

\section{Summary and outlook}
\label{sec:SUM}

In this review we have discussed quantum (or zero temperature)
phase transitions in electronic systems. After a pedagogical
introduction on the similarities and differences between classical thermal
and quantum phase transitions we have considered three specific examples
in some detail: quantum phase transitions in the
spherical model, magnetic quantum phase transitions of itinerant electrons,
and some aspects of the disorder-driven metal-insulator transition.
In this final chapter we want to summarize the results
from a common perspective and discuss some of the remaining open questions
as well as future research directions.

The theoretical description of a particular phase transition occurring
in nature usually starts with the identification of the relevant
variables, the most important one being the order parameter. To proceed
further, analytical investigations often follow the Landau-Ginzburg-Wilson
philosophy, i.e., all degrees of freedom other than the order parameter
fluctuations are integrated out, resulting in an effective theory
in terms of the order parameter only. If a rigorous analytical derivation
is complicated, the Landau-Ginzburg-Wilson theory is sometimes guessed
based on general symmetry considerations. (This is true in particular
when constructing toy models like the quantum spherical model.)
However, our work on the ferromagnetic
quantum phase transition has shown that great care has to be taken
in such an approach. At zero temperature an electronic system usually
contains many non-critical soft modes in addition to the critical
order parameter
fluctuations. If these soft modes couple to the order parameter
integrating them out leads to a singular behavior in the
Landau-Ginzburg-Wilson theory.
This mechanism is not restricted to zero temperature. Any soft mode
coupling to the order parameter can -- when integrated out -- produce
singularities in the resulting Landau-Ginzburg-Wilson theory.
However, in electronic
systems the number of soft modes at zero temperature is much higher than
at finite temperatures.

In the case of itinerant ferromagnets we have shown that these singularities
result either in unusual non-mean-field scaling behavior at the
quantum phase transition or in the quantum phase transition being of first
order (as is the case experimentally for the transition in MnSi).
Within the singular Landau-Ginzburg-Wilson theory it is very hard to
find out what scenario is realized for what microscopic parameters since
explicit calculations are next to impossible. A much more promising
though technically more challenging approach consists in {\em not}
integrating out all degrees of freedom other than the order parameter. Instead,
the effective field theory should treat all {\em soft} modes
in the system on the
same footing. We will explore the merits of such an approach in the future.

Another potential problem of the standard approach is caused by the fact
that it is based on perturbation theory: First, integrating out the
microscopic degrees of freedom in the derivation of the Landau-Ginzburg-Wilson
theory can usually be done only perturbatively. Second, the analysis of
the effective theory is done by means of the perturbative renormalization
group. Now, if the free energy landscape in configuration space contains
many, nearly degenerate minima separated by large barriers, as for instance
in the presence of random interactions or random fields, a
straight-forward perturbative approach may miss part of the physics.
We have shown this explicitly for the antiferromagnetic quantum phase
transition of weakly disordered itinerant electrons.
By approximately taking into account the non-perturbative degrees
of freedom connected with the many nearly degenerate free energy minima
we have derived a correction to the effective action of the standard
perturbative approach. This correction takes the form of static annealed
disorder. Including the new term into the renormalization group
destabilizes the conventional
critical behavior. However, within our theory the ultimate fate of
the transition could not be determined. Further work will be necessary
to decide between the different possibilities, viz., a complete destruction of
the antiferromagnetic long-range order or an unconventional transition
that may be characterized by an infinite-disorder fixed point and activated
scaling.

It should be emphasized here that we have also shown that including the
non-perturbative degrees of freedom connected with the many-valley structure
of the free energy in the presence of disorder does {\em not always} destroy
the conventional critical behavior. For itinerant quantum ferromagnets,
for instance, the long-range interaction between the order parameter fluctuations
stabilizes the conventional critical fixed point.
In principle, the method we used to include rare regions into magnetic
quantum phase transitions of disordered itinerant electrons can be
applied to any quantum phase transition in the presence of quenched disorder.
In the case of the disorder-driven metal-insulator transition we found
that magnetic rare regions (which physically correspond to local magnetic
moments) can lead to a new mechanism and a new universality class for the
transition. In the simplified model investigated so far (which neglects all
interaction effects beyond local moments) no transition was found in two
dimensions. An obvious question would be whether this conclusions changes
for a more realistic model.

An alternative way to go
beyond perturbation theory are numerical simulations. We have chosen this
approach for an investigation of the metal-insulator transition in the presence
of both interactions and quenched disorder. We have developed an efficient numerical method, the Hartree-Fock based
diagonalization method which is related to the configuration interaction
approach used in quantum chemistry. It allows us to calculate
with high accuracy the low-energy
properties of disordered interacting electrons for moderately large systems.
So far, we have mostly obtained qualitative results concerning the influence
of the electron-electron interactions on the transport properties: For
systems of spinless fermions the interactions reduce the d.c.\ conductance
for
weak disorder (in agreement with perturbative results) while they enhance
the d.c.\ conductance for strong disorder.
In future work we plan to extend this
work to a systematic finite-size scaling study which will lead to
quantitative results about the metal-insulator transition.
Moreover, we are working on including the spin degrees of freedom into
our numerical scheme. This is of particular importance since the spin
degrees of freedom have proven to be crucial for the recently
discovered metal-insulator transition in two dimensions.

\vspace*{0.25cm} \baselineskip=10pt{\small \noindent
The work presented here would have been impossible without the
contributions of many friends and colleagues. I particularly
thank my collaborators on the magnetic transitions,
D. Belitz, T. R.  Kirkpatrick, R. Narajanan and A. Millis. Equal thanks go
to M. Schreiber and F. Epperlein for the collaboration
on the numerical part of the work.
I greatly appreciated discussions with Alexei Efros, Ferdinand Evers,
Roger Haydock, Svetlana Kilina, Bernhard Kramer,
Arnulf M\"obius, Christian Pfleiderer,
Michael Pollak, Rudolf R\"omer, Achim Rosch,
Walter Schirmacher, Jorge Talamantes, John Toner,
Matthias Vojta, Dietmar Weinmann,
and Isa Zarekeshev.

This work was supported in part by the DFG  (Vo659/1, SFB 393/C2) and by
the NSF (DMR-95-10185, DMR-98-70597). The author is also grateful to the
Aspen Center for Physics, where part of the work has been performed.}

\end{document}